\documentclass[a4paper,11pt]{article}

\usepackage[a4paper,left=2.73cm,right=2.7cm,top=3cm,bottom=3.5cm]{geometry}

\usepackage{graphicx}
\usepackage{epsfig}
\usepackage{amssymb}
\usepackage{amsfonts}
			\usepackage{dsfont}															
			\newcommand{\mathset}{\mathds}							
\usepackage{amsmath,euscript,array,mathrsfs}
\usepackage{bbold}
\usepackage{epsf}
\usepackage{slashed}
\usepackage{comment}
\usepackage{colortbl}
\usepackage{cite}

\usepackage{caption,subcaption,wrapfig}
\usepackage[colorlinks=true,linktocpage=true,linkcolor=blue,citecolor=blue]{hyperref}

\usepackage{tikz}
\usetikzlibrary{decorations.markings,decorations.pathmorphing}

\newcommand{\be}{\begin{equation}}
\newcommand{\ee}{\end{equation}}
\newcommand{\beqs}{\begin{eqnarray}}
\newcommand{\eeqs}{\end{eqnarray}}
\newcommand{\parent}[1]{\left(#1\right)}
\newcommand{\mt}[1]{\textrm{\tiny #1}}
\newcommand{\Sec}[1]{Sec.~\ref{#1}}
\newcommand{\fig}[1]{Fig.~\ref{#1}}

\newcommand{\alignedeq}[1]{\begin{equation}
\begin{aligned}
#1
\end{aligned}\end{equation}
}

\newcommand{\Tr}{{\rm Tr}}

\newcommand{\dd}{\mathrm{d}}

\def\r{\rho}

\def\nc{N_\mt{c}}
\def\nf{N_\mt{f}}
\def\nc{N_\mt{c}}
\def\ns{N_\mt{s}}
\def\xs{x_\mt{s}}
\def\xf{x_\mt{f}}
\def\xc{x_\mt{s}^\mt{c}}

\begin{document}

 \begin{titlepage}

\thispagestyle{empty}

\begin{flushright}
\hfill{ICCUB-21-008}
\end{flushright}

\vspace{40pt}  
	 
\begin{center}

{\huge \textbf{Multiple Mass Hierarchies\\[3mm] from \\[7mm] Complex Fixed Point Collisions}}

\vspace{30pt}
		
{\large \bf Ant\'on F. Faedo,$^{1,\,2}$   Carlos Hoyos,$^{1,\,2}$   \\ [1mm]
David Mateos$^{3,\,4}$  and Javier G. Subils$^{3}$}

\vspace{25pt}

{\normalsize  $^{1}$ Departamento de F\'{i}sica, Universidad de Oviedo, \\ Federico Garc\'ia Lorca 18, ES-33007, Oviedo, Spain.}\\
\vspace{15pt}
{ $^{2}$Instituto Universitario de Ciencias y Tecnolog\'{\i}as Espaciales de Asturias (ICTEA), \\ Calle de la Independencia 13, ES-33004, Oviedo, Spain.}\\
\vspace{15pt}
{ $^{3}$Departament de F\'\i sica Qu\'antica i Astrof\'\i sica and Institut de Ci\`encies del Cosmos (ICC),\\  Universitat de Barcelona, Mart\'\i\  i Franqu\`es 1, ES-08028, Barcelona, Spain.}\\
\vspace{15pt}
{ $^{4}$Instituci\'o Catalana de Recerca i Estudis Avan\c cats (ICREA), \\ Passeig Llu\'\i s Companys 23, ES-08010, Barcelona, Spain.\\ 
}

\vspace{40pt}
				
\abstract{A pair of complex-conjugate fixed points that lie close to the real axis generates a large mass hierarchy in the real renormalization group flow that passes in between them. We show that pairs of complex fixed points that are close to the real axis and to one another generate multiple hierarchies, some of which can be parametrically enhanced. We illustrate this effect at weak coupling with field-theory examples, and at strong coupling using holography. We also construct complex flows between complex fixed points, including flows that violate the $c$-theorem.} 

\end{center}

\end{titlepage}

\tableofcontents

\hrulefill
\vspace{10pt}

\section{Introduction}
\label{intro}
Our fundamental description of the Universe is filled with hierarchies. The electron mass is 6 orders of magnitude smaller than the top-quark mass. Neutrino masses are even smaller by several orders of magnitude. The mixings in the quark sector, i.e.~the entries in the Cabibbo--Kobayashi--Maskawa matrix, span 3 orders of magnitude. The Higgs mass is 17 orders of magnitude smaller than the Planck mass. The dark energy density driving the accelerated expansion of the Universe  differs from the Planck scale by 123 orders of magnitude. This raises the question of what kind of Wilsonian dynamics may give rise to such multiple hierarchies or, somewhat poetically, to a hierarchy of hierarchies. 

\begin{figure}[t]
\begin{center}
\includegraphics[width=\textwidth]{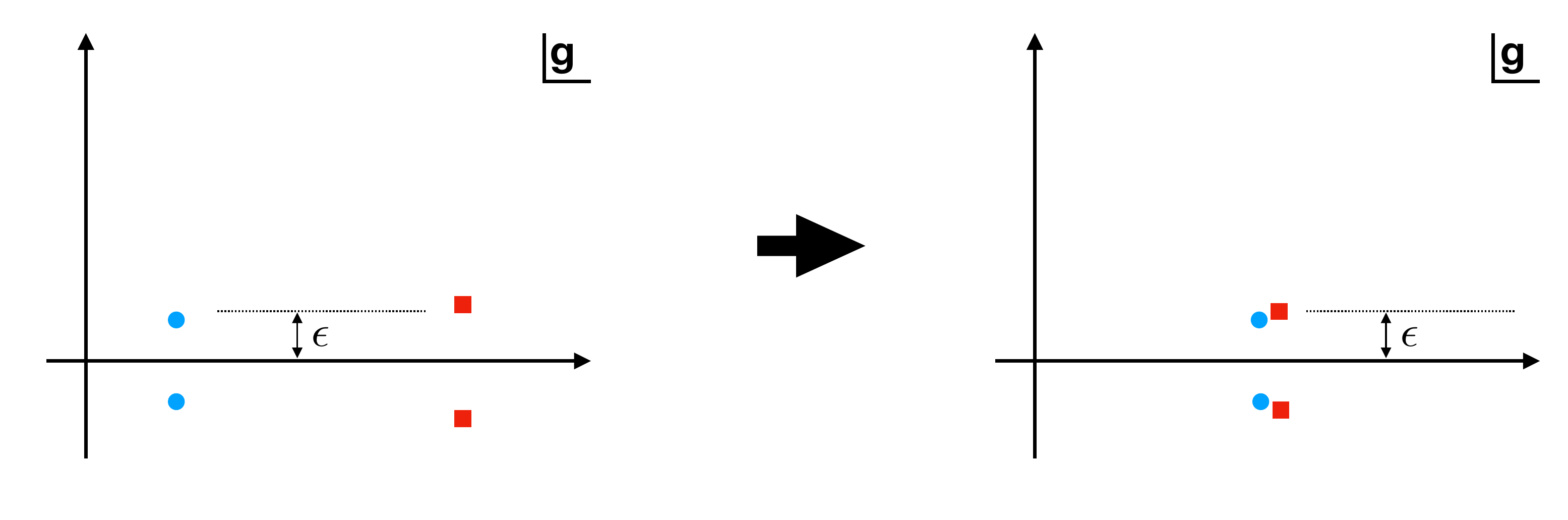} 
\caption{\small (Left) Two widely separated complex-conjugate pairs of cFPs close to the real axis. (Right) As a parameter in the theory is varied the two pairs can approach each other and ``collide''.}
\label{fig:collision}
\end{center}
\end{figure}

It is well known that pair of complex fixed points (cFPs) close to the real axis can generate a large hierarchy \cite{Kaplan:2009kr,Gorbenko:2018ncu,Gorbenko:2018dtm}. In \fig{fig:collision}(left) each of the two widely separated pairs (the blue circles and the red squares) generates one independent such hierarchy of approximately the same size. The goal of this paper is to show that pairs of cFPs that are close to the real axis and to one another, as in \fig{fig:collision}(right), generate multiple hierarchies of varying sizes, some of which are parametrically enhanced with respect to the case of isolated pairs. 

It is by now clear that the existence of FPs at complex values of the couplings is a generic feature in a large class of condensed-matter and high-energy field-theory models, at least at the perturbative level where the $\beta$-functions are polynomials in the couplings. Examples include systems with weak first-order transitions \cite{Halperin:1973jh,Nienhuis:1979mb,Nauenberg:1980nv,Senthil:2004,Herbut:2014lfa,Nahum:2015jya,Wang:2017txt,Serna:2018tct,Gorbenko:2018dtm,Ihrig:2019kfv}, six-dimensional $O(N)$ models \cite{Fei:2014xta,Gracey:2018khg} and gauge theories close to the boundary of the conformal window \cite{Appelquist:1988sr,Kubota:2001kk,Kaveh:2004qa,Herbut:2016ide,Gies:2005as,Pomoni:2008de,Kaplan:2009kr,Antipin:2012kc,Hansen:2017pwe}.

A well understood mechanism that gives rise to these cFPs is the annihilation of pairs of real FPs. As we vary the parameters of the model, such as the number of fields or the number of dimensions, two fixed points on the real axis can approach each other, merge and migrate to the complex plane in the form of complex-conjugate pairs. This phenomenon of fixed point annihilation (FPA) was argued in \cite{Kaplan:2009kr} to be responsible for the walking behavior of some gauge theories and in \cite{Gorbenko:2018ncu,Gorbenko:2018dtm} to give rise to weak first-order transitions in condensed-matter models. 

A large hierarchy of scales is generated when the pair of cFPs is close to the real axis and the renormalization group (RG) flow passes in between them.  ``Proximity'' to the real axis is not always a precise notion, since $\beta$-functions or couplings are not directly observable quantities. Thus, sometimes a better characterization of the hierarchy may be obtained through the concept of complex conformal field theories (cCFTs) that were conjectured in \cite{Gorbenko:2018ncu, Gorbenko:2018dtm} to exist at any given cFP. In such cCFTs, which occur in complex-conjugate pairs, there is a well defined notion of the dimension of an operator, completely analogous to that in usual CFTs. However, since cCFTs are non-unitary, these dimensions are in general complex numbers. Let $\Delta$ denote the dimension at one of the cCFTs  of the operator driving the flow. Then there is a large hierarchy of scales if its imaginary part verifies $|\operatorname{Im}{\Delta}|\sim \epsilon \ll 1$, with $\epsilon$ a tunable parameter. Specifically, the scales $\mu_{\text{\tiny UV}}$ and $\mu_{\text{\tiny IR}}$ associated to the hierarchy obey Miransky scaling \cite{Miransky:1984ef} 
\be
\label{eq:Miransky}
\log \frac{\mu_{\text{\tiny UV}}}{\mu_{\text{\tiny IR}}} \simeq 
\frac{2\pi}{|\operatorname{Im}{\Delta}|} \,.
\ee
This illustrates how cCFTs are useful to characterize physical properties of real flows. In the case of isolated pairs of cFPs the right-hand side above typically scales as $1/\epsilon$ when expressed in terms of the imaginary part of the value of the complex coupling at the cFP --- see  \fig{fig:collision}.

The mechanism we just described only requires the existence of one pair of real FPs and a suitable parameter so that by tuning it they annihilate into a complex pair. However, cFPs are frequent and can appear in groups of more than one pair, for instance when considering perturbative $\beta$-functions at high-loop orders, which give rise to high-rank polynomials in the couplings, or in models with several couplings. In this paper we will study the  interplay between several pairs of cFPs. We will pay  particular attention to the resulting modifications of the usual scaling \eqref{eq:Miransky} when two or more pairs of cFPs ``collide'' in the complex plane as some parameter is varied. We emphasize that our goal is not to explain any of the specific hierarchies of the real world but to provide a proof of concept that colliding cFPs can  generate multiple hierarchies with modified scalings. 

The presence of additional FPs, complex or real, can influence the Miransky scaling at different levels.\footnote{The influence we have in mind goes beyond the polynomial corrections to the exponential hierarchy \eqref{eq:Miransky} computed in \cite{Jarvinen:2010ks, Braun:2010qs}.} The mildest adjustment would be to change the coefficient $2\pi$ in \eqref{eq:Miransky} so that the hierarchy is still controlled by (the inverse of) $|\operatorname{Im}{\Delta}|$ albeit with a different slope. A more profound modification would be to replace the physical quantity that governs the scaling. This must certainly be the case in the singular limit $|\operatorname{Im}{\Delta}|=0$, which is realised when two cFPs merge. In this case the corresponding $\beta$-function has a double zero and, as a consequence, the operator driving the flow is marginal at the cCFTs and hence  $\Delta= d\in\mathset{R}$. As we will see, this is typically accompanied by an enhancement of the hierarchy from ${\mu_{\text{\tiny UV}}}/ {\mu_{\text{\tiny IR}}} \sim \exp(1/\epsilon)$ to ${\mu_{\text{\tiny UV}}}/ {\mu_{\text{\tiny IR}}} \sim \exp(1/\epsilon^3)$. 

In the following we will study examples of these situations in models possessing several conjugate pairs of cFPs and possibly additional real ones. Along the way we will uncover interesting properties of the RG flows involving cFPs. We will analyze cCFTs at weak and at strong coupling. In the first case we will use perturbation theory to study both toy-model $\beta$-functions and fully-fledged quantum field theory examples. In the second case we will use holography and the recently proposed gravitational duals of cFPs \cite{Faedo:2019nxw}.

\section{Weak coupling realization}
\label{sec:weak}

In this section we study models that admit several pairs of cFPs in the perturbative regime. We first discuss the general properties by means of simple toy-model $\beta$-functions and then show explicit field-theoretic examples admitting multiple pairs of cFPs that can be made to coincide. 

\subsection{Toy models for the $\beta$-function}
\label{sec:toy}

We are interested in the interplay between several pairs of cFPs and how they influence the physics of Miransky scaling. The fact that Eq. \eqref{eq:Miransky} must be modified if another pair of fixed points comes near the original one can be seen as follows. Consider for simplicity a model with a single running coupling for some scalar operator. When two real FPs meet, just before migrating to the complex plane, one has a double FP. Since the first derivative of the $\beta$-function vanishes at that point, the operator becomes marginal when the two FPs merge. 

In the very same way, if we  tune the model so as to make two pairs of complex-conjugate FPs coincide, the operator driving the flow would be marginal with respect to the  cCFT defined at that precise point, again because the derivative of the $\beta$-function must vanish. Since marginal operators have purely real dimension 
$\Delta=d$, Eq.~\eqref{eq:Miransky} cannot be correct in the limit in which two cFPs collide. By continuity, just before the collision the scaling has to change. Furthermore, it has to be enhanced, since the hierarchy of scales is enlarged as $\operatorname{Im}{\Delta}\to0$. A natural question is then what replaces $|\operatorname{Im}{\Delta}|$ as the quantity controlling this hierarchy.

\begin{figure}[t]
\begin{center}
		\begin{subfigure}{0.41\textwidth}
			\includegraphics[width=\textwidth]{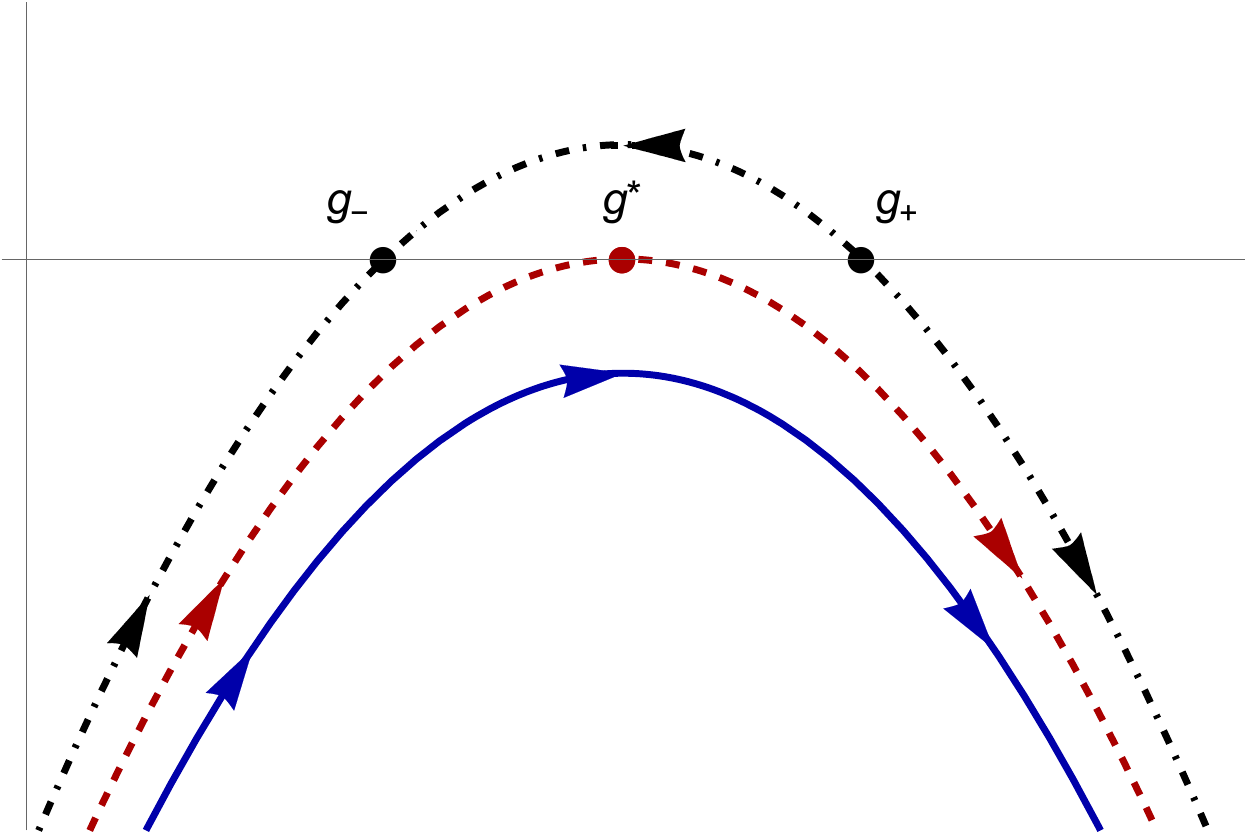} 
			\put(-182,125){\Large $\beta$}
			\put(3,80){\Large $g$}
		\end{subfigure}\hspace{10ex}
		\begin{subfigure}{.41\textwidth}
			\includegraphics[width=\textwidth]{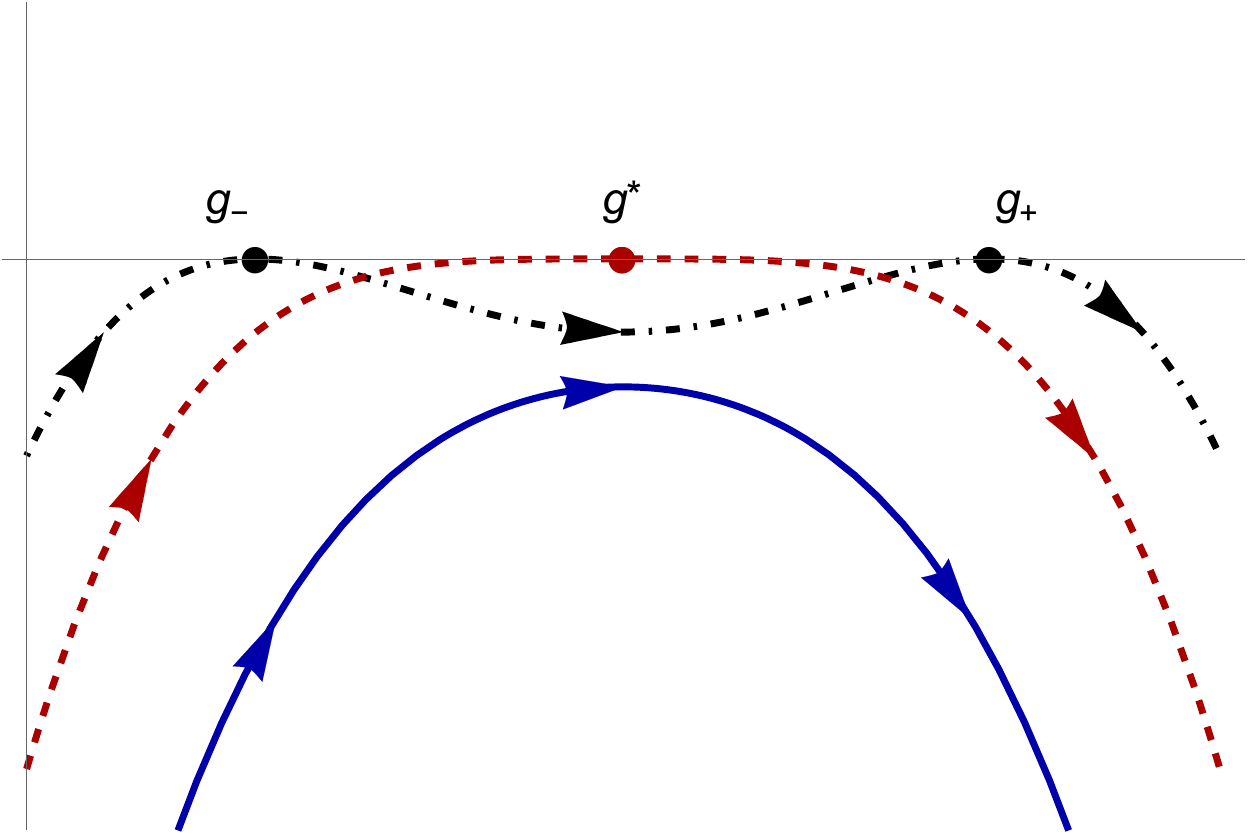} 
			\put(-182,125){\Large $\beta$}
			\put(3,80){\Large $g$}
		\end{subfigure}
		\vspace{4mm}
		\caption{\small The $\beta$-functions \eqref{eq:betasimp} (left) and \eqref{eq:betadouble} (right). In both cases two FPs (dashed-dotted, black curve) merge (dashed, red curve) an eventually disappear into the complex plane (solid, blue curve). On the left panel these are single FPs. On the right panel they are double FPs and, as a consequence,  the operator corresponding to $g$ is marginal both at the UV and at the IR FPs present on the black curve. Marginality is inherited by the double cFPs that result after the annihilation. Notice that a double FP cannot annihilate with a single FP.}
		\label{fig:annihilation}
	\end{center}
\end{figure}

Following the logic of FPA described in \Sec{intro}, this process can be seen equivalently as the annihilation of two double, real FPs that become a pair of double cFPs, where by a double FP we mean a double-zero of the $\beta$-function. Therefore, the question can be rephrased as what is the physics of double real FPA, as shown in the right panel of \fig{fig:annihilation}. 

Before analyzing how double FPs annihilate, let us review the usual mechanism of single FPA described in \cite{Kaplan:2009kr}. Consider a $\beta$-function that near $g=g^*$ takes the simple form
\begin{equation}\label{eq:betasimp}
\beta\left(g\right)=-\epsilon^2-\left(g-g^*\right)^2+O\left((g-g^*)^3\right)\,.
\end{equation}
Notice in particular the absence of a linear term, meaning that there is a maximum at $g=g^*$. The tunable constant $\epsilon^2$ is a function of the parameters that specify the theory such as the number of fields, the spacetime dimension, etc. The structure of fixed points depends heavily on the sign of this constant. For $\epsilon^2<0$ there are two real FPs located at $g_\pm=g^*\pm |\epsilon|$. The operator is relevant at $g_+$ and irrelevant at $g_-$ and there is an RG flow connecting the associated CFTs. 

As $\epsilon^2\to 0$, the dimension of the operator increases at $g_+$ and decreases at $g_-$. Consequently, it becomes marginal when $g_+=g_-$, corresponding precisely to $\epsilon^2=0$. This is the value at which the fixed points annihilate, since for $\epsilon^2>0$ there are no real zeroes of the $\beta$-function and therefore no real FPs. These three possibilities are shown in the left panel of \fig{fig:annihilation}.

When $\epsilon^2>0$ the $\beta$-function still vanishes at the complex values of the coupling \mbox{$g=g^*\pm i\,\epsilon$}. The proposal of \cite{Gorbenko:2018ncu, Gorbenko:2018dtm} is that a certain type of non-unitary conformal field theory, dubbed complex CFT, can be defined at each of the cFPs. Since the FPs come in complex-conjugate pairs, so do the cCFTs. For the example described by \eqref{eq:betasimp} the dimensions of the operators at the corresponding cCFTs are 
$\Delta=d\mp2 i \epsilon+O\left(\epsilon^2\right)$. This is a general feature of cCFTs: if one cCFT in the pair possesses an operator of dimension $\Delta$, the complex-conjugate CFT possesses an  operator of dimension $\Delta^*$.

Consider now the real flow corresponding to $\epsilon^2>0$, as described by the blue curve in \fig{fig:annihilation}, which passes in between the cFPs.  Direct integration of \eqref{eq:betasimp} up to quadratic order shows that, when $\epsilon\ll 1$ and thus the cFPs are close to the real axis, there is a large hierarchy of scales given by
\begin{equation}
\log \frac{\mu_{\text{\tiny UV}}}{\mu_{\text{\tiny IR}}}=\int^{g_{\text{\tiny UV}}}_{g_{\text{\tiny IR}}}\frac{\dd g}{\beta(g)}\simeq\frac{\pi}{\epsilon} \,.
\end{equation}
Written in terms of $|\operatorname{Im}{\Delta}|$ this gives precisely \eqref{eq:Miransky}. The picture that emerges from this mechanism of FPA is that RG flows passing in between cFPs are slowed down as long as these are close to the real axis as quantified by $|\operatorname{Im}{\Delta}|$.

Let us now see how the annihilation of double FPs takes place for a single coupling. Suppose now that, in the vicinity of $g=g^*$, the $\beta$-function reads 
\begin{equation}\label{eq:betadouble}
\beta\left(g\right)=-\left[\left(g-g^*\right)^2+\epsilon^2\right]^2+O\left((g-g^*)^5\right)\,.
\end{equation}
There are again three different possibilities according to the value of the independent term. If $\epsilon^2 < 0$ there are two zeroes at $g_\pm=g^*\pm|\epsilon|$, each of them having multiplicity two. As we have seen, the dimension of the operator at double FPs is $\Delta=d$ in $d$ dimensions, i.e.~the operator is marginal.\footnote{More precisely, it would need to be marginally relevant at the UV FP and marginally irrelevant at the IR FP for a flow between these FPs to exist.} When $\epsilon^2=0$ the FPs merge into a quadruple one. Increasing the parameter further to $\epsilon^2>0$ they migrate into the complex plane and are located in pairs at $g=g^*\pm i\,\epsilon$. Notice that in all this process the operator is marginal at every FP, whether this is real or complex, since the $\beta$-function is at least quadratic at each of them. This is captured by the right panel of \fig{fig:annihilation}.

After the annihilation, there is a large separation of scales when the double cFPs are close to the real axis and to one another given by
\begin{equation}
\log \frac{\mu_{\text{\tiny UV}}}{\mu_{\text{\tiny IR}}}=\int^{g_{\text{\tiny UV}}}_{g_{\text{\tiny IR}}}\frac{\dd g}{\beta(g)}\simeq\frac{\pi}{4\epsilon^3} \,,
\end{equation}
with $\epsilon\ll1$. The $1/\epsilon^3$ scaling on the right-hand side constitutes an enhancement with respect to the usual $1/\epsilon$ behaviour,  meaning that the hierarchy of scales is much larger in this case for a  fixed, small $\epsilon$. This may not seem an entirely meaningful comparison, since $\epsilon$ is simply the parameter that controls the imaginary part of the cFP and as we know $\beta$-functions are not directly  physical observables. However, as we will see in explicit examples, the role of $\epsilon$ is often played by physical  parameters such as the number of fields in the theory, so the enhancement above is a meaningful statement.  

Another crucial difference with the previous situation is that now the operator is also marginal at the cCFTs, so $|\operatorname{Im}{\Delta}|=0$ and the hierarchy cannot be expressed in terms of $\operatorname{Im}{\Delta}$. One would need to find the correct scheme-independent quantity defined on the cCFTs which is $O\left(\epsilon^3\right)$ and could be used to quantify the scaling in terms of purely CFT data. 

This observation seems to be particular to the case of marginal operators, that is, when two cFPs coincide, so in order to shed light on this question let us see in detail how the situation changes as the cFPs get closer. In general, two cFPs can approach each other along an arbitrary direction on the complex plane. For simplicity, we begin by considering the case where they approach each other along a horizontal line. Although this may seem fairly non-generic, this situation actually captures the essential features that we wish to highlight. We therefore assume that the two cFPs have the same imaginary part $\epsilon$ and that their real parts differ by $2\delta$. We are interested in keeping $\epsilon$ small but fixed and varying 
$\delta$. 

We thus consider a $\beta$-function of the form
\begin{equation}\label{eq:betaeqim}
\beta=-\left[\left(g-g^*+\delta\right)^2+\epsilon^2\right]\left[\left(g-g^*-\delta\right)^2+\epsilon^2\right]\,,
\end{equation}
with $\epsilon^2>0$. This reduces to \eqref{eq:betadouble} for $\delta\to0$. We can think of this parameter as external and tunable in the same way as $\epsilon$. There are two pairs of complex-conjugate FPs located at
\begin{equation}
g=g^*\pm\delta\pm i\,\epsilon\,.
\end{equation}
There is a symmetry $\delta\leftrightarrow-\delta$, so from now on we will consider $\delta>0$ and study in detail what happens around one of the FPs as the second one comes closer. The dimension of the operator associated to $g$ at the different cFPs is
\begin{equation}\label{eq:dimension}
\Delta=d\pm8\,i\,\delta^2\epsilon\pm8\,\delta\epsilon^2\,,
\end{equation}
which of course becomes marginal when the fixed points coincide, $\delta\to 0$.

We expect walking behaviour when the FPs are close to the real axis, that is, in the vicinity of a maximum. The extrema are located at
\begin{equation}\label{eq:extrema}
g=g^*\,,\qquad\qquad\qquad g=\widetilde{g}_\pm^*=g^*\pm\sqrt{\delta^2-\epsilon^2}\,.
\end{equation}
It is then clear that we have to distinguish three cases depending on the hierarchy between the parameters $\epsilon$ and $\delta$. From \eqref{eq:dimension} it is easy to see that $\delta/\epsilon=|\operatorname{Im}\Delta/\operatorname{Re}(d-\Delta)|$, so the analysis that follows can be equivalently stated in terms of the real and imaginary parts of the scaling dimension at the cFPs. 

\begin{itemize}

\item \textit{\textbf{Case A:} $\delta\gg\epsilon$}.

In this case the two pairs of cFPs are widely separated and we do not expect a dramatic influence between them. The first solution in \eqref{eq:extrema} is a minimum while the other two are maxima located symmetrically around it. By scaling the coupling it can be seen that the $\beta$-function around any of the maxima can be put into the form
\begin{align}
\beta\left(g\right)&=-16\,\delta^2\,\epsilon^2\left(\delta^2-\epsilon^2\right)-\left(g-\widetilde{g}_\pm^*\right)^2+O\left((g-\widetilde{g}_\pm^*)^3\right)\nonumber\\[2mm]
&\simeq-16\,\delta^4\,\epsilon^2-\left(g-\widetilde{g}_\pm^*\right)^2+O\left((g-\widetilde{g}_\pm^*)^3\right)\,.
\end{align}
where we have discarded terms of $O\left(\epsilon^2/\delta^2\right)$.

This is formally as \eqref{eq:betasimp}, so assuming $\delta^2\epsilon\ll1$ the same argument predicts the hierarchy of scales 
\begin{equation}
\log \frac{\mu_{\text{\tiny UV}}}{\mu_{\text{\tiny IR}}}\simeq\frac{\pi}{4\delta^2\epsilon}=\frac{2\pi}{|\operatorname{Im}{\Delta}|}\,,
\end{equation}
where we have used the dimension given by \eqref{eq:dimension}. Thus, as expected, the second pair of cFPs has little impact and the usual Miransky scaling \eqref{eq:Miransky}, ${\mu_{\text{\tiny UV}}}/ {\mu_{\text{\tiny IR}}} \sim \exp(1/\epsilon)$, is reproduced. 

Notice that, when the RG flow passes in between the second pair of cFPs, there will be an analogous walking region  controlled by the imaginary part of the scaling dimension at the corresponding cCFTs. Indeed, as the distance between the cFPs decreases the walking regions start to superpose and the hierarchies add up. This can be seen from direct integration of the $\beta$-function, where we must keep terms up to quartic order, since  otherwise we would not detect the presence of the second pair of cFPs that enlarges the walking region. Moreover, we must ensure that the integration is performed on an interval $\left(g_{\text{\tiny IR}},g_{\text{\tiny UV}}\right)$ that includes both pairs. In this case the resulting scaling is 
\begin{equation}\label{eq:twice}
\log \frac{\mu_{\text{\tiny UV}}}{\mu_{\text{\tiny IR}}}\simeq\frac{2\pi}{|\operatorname{Im}{\Delta}|}+\frac{2\pi}{|\operatorname{Im}{\Delta}|}=\frac{4\pi}{|\operatorname{Im}{\Delta}|}
\,.
\end{equation}
We see that, for this simple model in which the imaginary part of $\Delta$ is the same at both pairs cFPs, the extent  of the walking region (on a log scale) is twice the usual one. The exact expression in terms of complex CFT data would differ for more complicated models, but it would still be of $O\left(1/\epsilon\right)$. With this example we have seen that the presence of nearby complex fixed points can affect the size of the walking region. Had we not known the existence of the second pair, we could have inferred it from the enlargement of the hierarchy. On the other hand, the fact that we have discarded terms of $O\left(\epsilon^2/\delta^2\right)$ shows that we should expect some effect to take place as both parameters become of the same order. 

\item \textit{\textbf{Case B:} $\delta\sim\epsilon$}.

As the distance between the pairs of cFPs $\delta$ decreases new terms start to compete with the leading one. When both $\epsilon$ and $\delta$ are of the same order, direct integration of the $\beta$-function \eqref{eq:betaeqim} gives the scaling 
\begin{equation}\label{eq:interpolate}
\log \frac{\mu_{\text{\tiny UV}}}{\mu_{\text{\tiny IR}}}\simeq\frac{\pi}{2\left(\delta^2\epsilon+\epsilon^3\right)}=\frac{4\pi}{|\operatorname{Im}{\Delta}|}\,\frac{\left(\operatorname{Im}{\Delta}\right)^2}{\left(\operatorname{Im}{\Delta}\right)^2+\left(\operatorname{Re}{\gamma}\right)^2}\,,
\end{equation}
where $\gamma=d-\Delta$ is the dimension of the coupling to the operator at the cFP. Again, this rewriting in terms of complex CFT data is model-dependent, but the crucial point is that, when $\delta\sim\epsilon$, both terms in the denominator of \eqref{eq:interpolate} are of the same order. As a result, the hierarchy is 
$O\left(1/\epsilon^3\right)$ instead of the usual $O\left(1/\epsilon\right)$ that we encountered in the previous case. Therefore, independently of the details of the model, when the second pair of cFPs is in the vicinity of the first one, so that their separation is comparable to their distance to the real axis, there is an enhancement of the walking region. This effect remains as the FPs become even closer and eventually coincide.

\item \textit{\textbf{Case C:} $\delta\ll\epsilon$}

For $\delta<\epsilon$ the first solution in \eqref{eq:extrema} is the unique maximum. Integration of the $\beta$-function in its vicinity produces the hierarchy 
\begin{equation}\label{eq:scalingC}
\log \frac{\mu_{\text{\tiny UV}}}{\mu_{\text{\tiny IR}}}\simeq\frac{\pi}{2\epsilon^3}=\frac{4\pi}{|\operatorname{Im}{\Delta}|}\,\left(\frac{\operatorname{Im}{\Delta}}{\operatorname{Re}{\gamma}}\right)^2\,,
\end{equation}
which corresponds to the second term in the denominator of \eqref{eq:interpolate}, which  is now dominant. We have thus discarded terms $O\left(\delta^2/\epsilon^2\right)$ smaller. This is the same scaling as the previous case, $O\left(1/\epsilon^3\right)$. The natural quantity in the cCFT controlling the hierarchy is again model-dependent. In the limit $\delta\to0$ both the real and imaginary parts of $\gamma$ vanish making the operator exactly marginal, so when the cFP is double one needs yet another quantity, probably related to a higher-point function. 

\end{itemize}

This example is just a proof of principle and, as we have been emphasizing, the exact expressions of the scaling in terms of cCFT data is model-dependent. We have studied in detail the case in which the cFPs approach each other along the real axis, but similar results are obtained if they approach along an arbitrary direction on the complex plane, the qualitative physics being sensitive only to the distance between the cFPs. The general case where there is an arbitrary number of complex fixed points at arbitrary positions is treated in Appendix \ref{sec:arbitrary}. This shows that the enhancement of the hierarchy from $O\left(1/\epsilon\right)$ to $O\left(1/\epsilon^3\right)$ when two pairs of cFPs are close to one another is robust.

\begin{figure}[t]
	\begin{center}
			\includegraphics[width=0.50\textwidth]{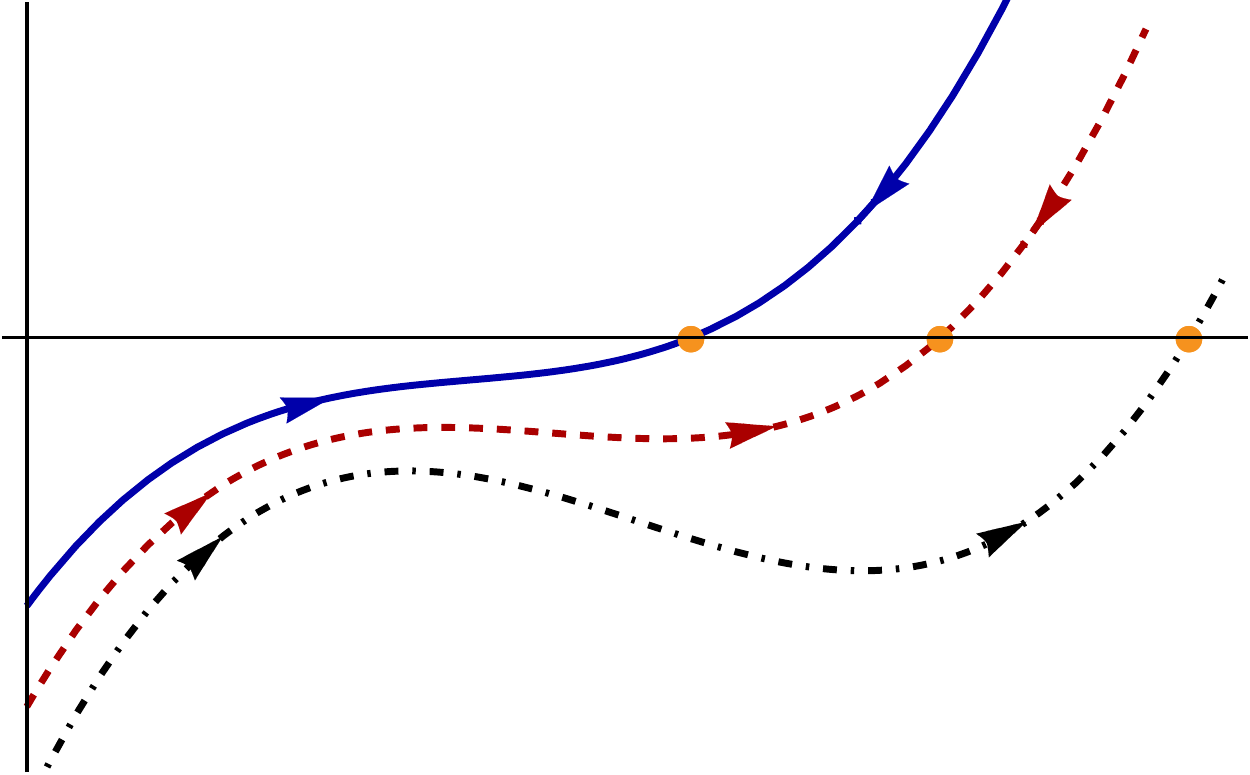} 
			\put(-228,130){$\beta$}
			\put(0,60){$g$}
		\caption{\small  Example of a $\beta$-function admitting a pair of cFPs and a real FP. The orange dot corresponds to the real fixed point at $g=g_*+\delta$. The real part of the cFPs is located roughly at the local maximum of the black curve. When the RG flow enters the vicinity of the maximum it slows down and starts to walk. As the real FP approaches the cFPs the maximum is lost and as a result the Miransky scaling disappears.}\label{fig:destruction}
	\end{center}
\end{figure}

We close this section discussing how the presence of a nearby real FP can spoil the Miransky scaling. Imagine a $\beta$-function that admits a pair of complex-conjugate cFPs at a distance $\epsilon\ll1$ from the real axis together with a real FP separated a distance $\delta$ from the cFPs, as in 
\begin{equation}\label{eq:spoil}
\beta=\left[\left(g-g^*\right)^2+\epsilon^2\right]\Big[g-\left(g^*+\delta\right)\Big]+O\left(g^4\right)\,.
\end{equation}
When the real FP is far from the cFPs, $\delta\gg1\gg\epsilon$, the RG flow passing in between the cFPs enjoys the usual walking with Miransky scaling \eqref{eq:Miransky} before reaching the real IR FP. However, if by tuning some parameter we can make the real FP approach the cFPs, so that $\delta\sim\epsilon$, then the  Miransky scaling disappears. The situation is depicted in Fig.~\ref{fig:destruction}, where we see that the presence of the real FP eventually destroys the maximum in the $\beta$-function which is needed for the Miransky scaling to exist. 

\subsection{Field theory examples}
\label{sec:FT}

In this section we provide explicit field theory examples having the general features discussed in the simple toy-models of the previous section. For this we need models admitting several pairs of cFPs. At the same time, in order to change the relative location of the FPs on the complex plane, they must possess enough tunable parameters. 
This is more easily achieved in models with several couplings, which moreover are phenomenologically more relevant and also have the advantage of admitting various pairs of cFPs at a lower-loop level. In addition, the interplay of the different couplings gives rise to additional interesting features such as the generation of multiple scales of different sizes. 

\subsubsection{A Yang--Mills example}
First we show that the model in \cite{Hansen:2017pwe, Benini:2019dfy} contains two pairs of complex-conjugate cFPs that  can be made to coincide. At the point where the pairs of cFPs merge, there is a marginal operator from the point of view of the cCFTs defined on them and, as a consequence, an enhancement of the size of the walking region in a sense that we will explain. 

The model consists of a Yang--Mills (YM) theory with SU($\nc$) gauge symmetry coupled to $\ns$ complex scalars $\phi_i$ together with $\nf$ Dirac fermions $\psi_a$, both of them in the fundamental representation of the gauge group. The Lagrangian takes the form
\begin{equation}
\mathcal{L}=-\frac12\Tr\, F^{\mu\nu}F_{\mu\nu}+\Tr\,\left(i\overline{\psi}  \slashed{D}\psi\right)+\Tr\,\left(D_\mu\phi^\dag D^\mu\phi\right)-h\,\Tr\,\left(\phi^\dag\phi\right)^2-f\left(\Tr\,\phi^\dag\phi\right)^2 \,,
\end{equation}
where sums over all the scalar and all the fermion fields are implicit. Besides the 't Hooft coupling $\lambda$,  there are couplings $h$ and $f$ for the complex scalars associated, respectively, to the single- and double-trace operators 
\be
\label{operators}
\mathcal{O}_h= \Tr\,\left(\phi^\dag\phi\right)^2 \,, \qquad 
\mathcal{O}_f = \left(\Tr\,\phi^\dag\phi\right)^2 \,.
\ee
We will use the simplified large-$N$ Veneziano-limit results of \cite{Benini:2019dfy}. In this approximation, the $\beta$-functions for the complete set of couplings read
\begin{align}\label{eq:betaBIS}
\beta_\lambda&=-\frac{22-\xs-4\xf}{3}\lambda^2+\frac23\left(4\xs+13\xf-34\right)\lambda^3\,,\nonumber\\[2mm]
\beta_h&=4(1+\xs)h^2-6\lambda\, h+\frac34\lambda^2\,,\\[2mm]
\beta_f&=4f^2+8(1+\xs)f h+12 \xs h^2-6\lambda f+\frac{3\xs}{4}\lambda^2\,,\nonumber
\end{align}
where we have defined the ratios  
\be
\label{ratiosNN}
\xf=\frac{\nf}{\nc}\,, \qquad \xs=\frac{\ns}{\nc} \,. 
\ee
Note that we are working at two-loop order in the gauge coupling and at one-loop order in the scalar couplings. This is consistent because the $\beta$-function for $\lambda$ at this order decouples from the rest. One consequence of this is that it can be solved independently. Moreover, working to two-loop order in $\lambda$ is necessary in order to identify, in addition to the Gaussian FP at $\lambda=0$, an interacting Banks--Zaks-type FP at a value 
\be
\label{lambda}
\lambda^*=\frac{\xs+4\xf-22}{68-26\xf-8\xs} \,.
\ee
It is important that, for a given value of $\xs$, this can be made arbitrarily small by tuning the number of fermions 
$\xf$. This feature is analogous to that of the FP close to the boundary of the Conformal Window in QCD, and it ensures that perturbation theory can be applied  reliably. 

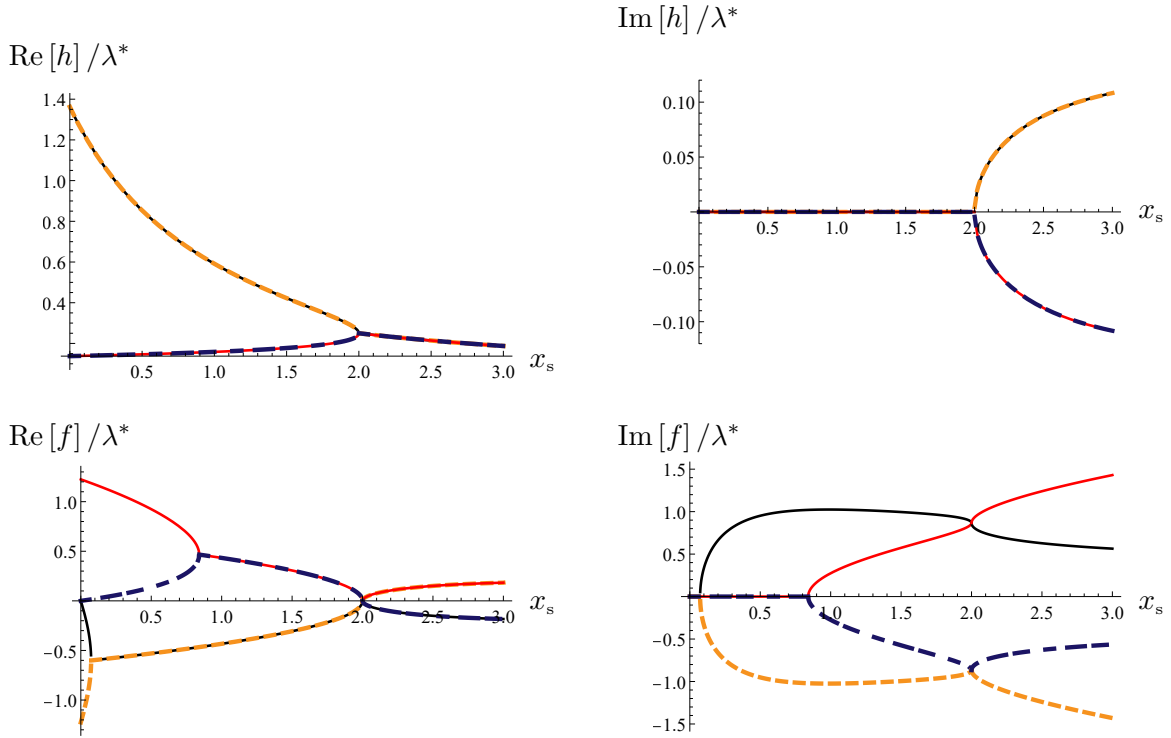
\begin{figure}[t]
\begin{center}
		\begin{subfigure}{0.40\textwidth}
			\includegraphics[width=\textwidth]{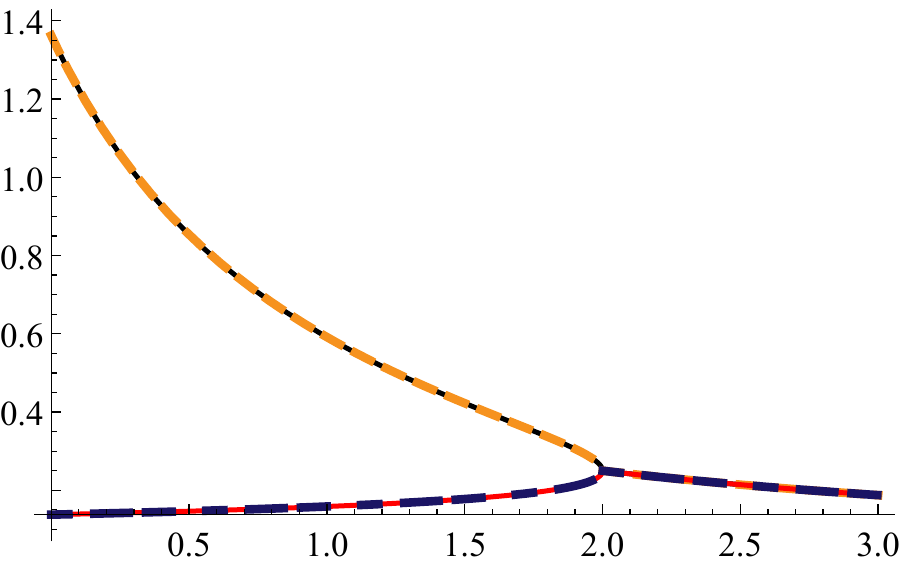} 
			\put(-190,120){${\rm Re}\left[h\right]/\lambda^*$}
			\put(5,6){$\xs$}
		\end{subfigure}\hspace{10ex}
		\begin{subfigure}{.40\textwidth}
			\includegraphics[width=\textwidth]{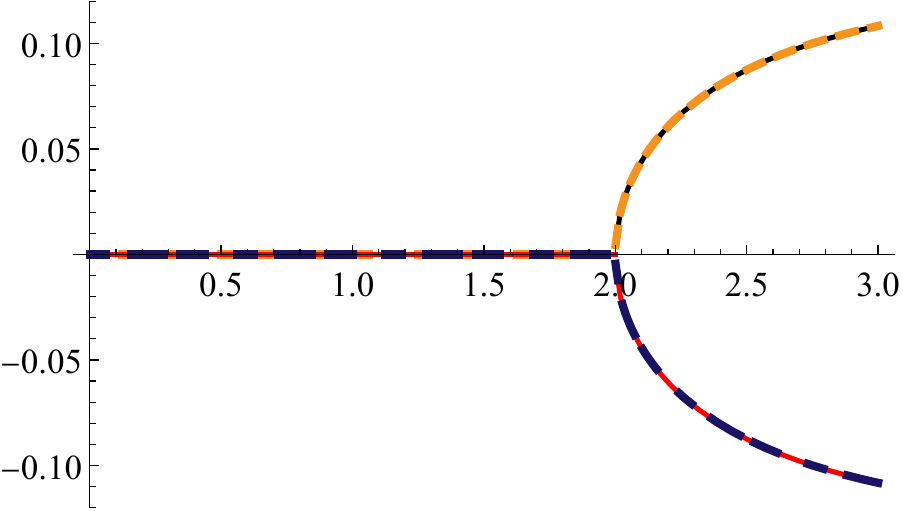} 
			\put(-190,120){${\rm Im}\left[h\right]/\lambda^*$}
			\put(5,48){$\xs$}
		\vspace{10mm}
		\end{subfigure}
		\begin{subfigure}{.40\textwidth}
			\includegraphics[width=\textwidth]{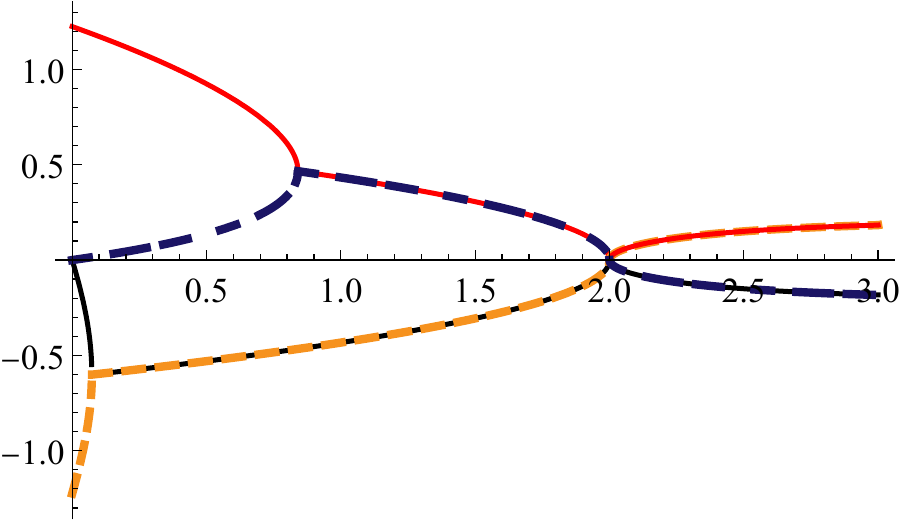} 
			\put(-190,110){${\rm Re}\left[f\right]/\lambda^*$}
			\put(5,48){$\xs$}
		\end{subfigure}\hspace{10ex}
	        \begin{subfigure}{.40\textwidth}
			\includegraphics[width=\textwidth]{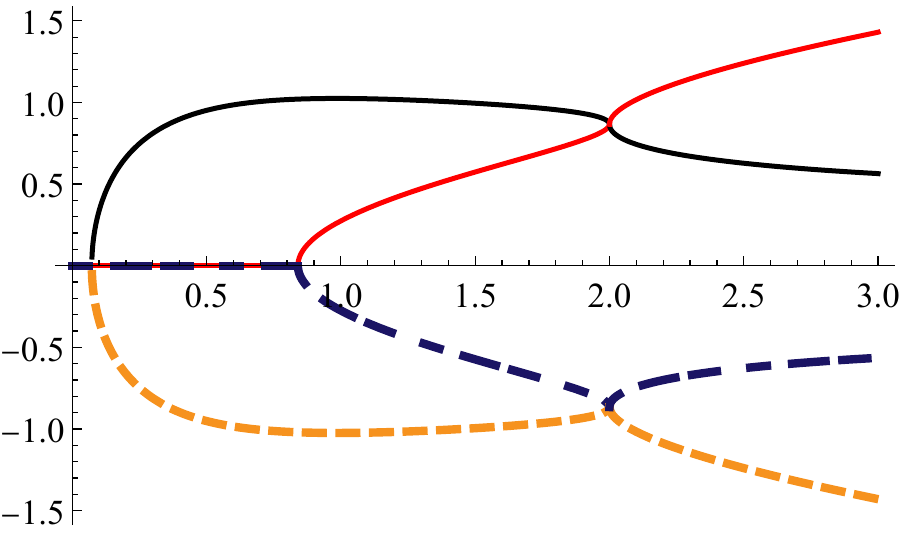} 
			\put(-190,110){${\rm Im}\left[f\right]/\lambda^*$}
			\put(5,48){$\xs$}
		\end{subfigure}
		\caption{\small Real and imaginary parts of the scalar couplings at the FPs as a function of the ratio between the number of scalars and colors defined in \eqref{ratiosNN}. At small values of $\xs$ there are four real FPs that annihilate in pairs and become complex at the two special values \eqref{crit1} and \eqref{crit2}. Above the latter value there are two pairs of conjugate cFPs (seen in the plot for the coupling ${\rm Im}[f]$) that coalesce precisely at $\xc=2$. Beyond that point we recover four cFPs and moreover $h$ also becomes complex.}
		\label{fig:lhf}
	\end{center}
\end{figure}

Once $\lambda$ is set equal to its FP value, the model admits four weakly-coupled IR FPs located at 
\begin{align}
{\rm FP}_1&=\left\{\lambda^*,\,h_{+}^*,\,f_{++}^*\right\}\qquad\qquad\qquad {\rm FP}_3=\left\{\lambda^*,\,h_{-}^*,\,f_{+-}^*\right\}\nonumber\\[2mm]
{\rm FP}_2&=\left\{\lambda^*,\,h_{+}^*,\,f_{-+}^*\right\}\qquad\qquad\qquad {\rm FP}_4=\left\{\lambda^*,\,h_{-}^*,\,f_{--}^*\right\}
\end{align}
where
\begin{equation}
h_{\pm}^*=\lambda^*\frac{3\pm4B}{4(1+\xs)}\,,\qquad\qquad f_{\pm+}^*=\lambda^*\left(-B\pm A_+\right)\,,\qquad\qquad f_{\pm-}^*=\lambda^*\left(+B\pm A_-\right)\,,
\end{equation}
with
\begin{equation}
\label{AB}
A_\pm=\frac{\sqrt{6-3\left(13\pm24B\right)\xs+3\xs^2-6\xs^3}}{4(1+\xs)}\,,\qquad\qquad\qquad B=\frac{\sqrt{6-3\xs}}{4}\,.
\end{equation}
We see that the location of the FPs is given in terms of two parameters: $\lambda^*$, which sets the overall scale, and $\xs$. For small enough values of $\xs$ all fixed points are real. By increasing this ratio it can be seen that the FPs annihilate in pairs and become complex. At a first critical value
\begin{equation}
\label{crit1}
\xs^{(1)}\simeq0.07309
\end{equation}
the quantity $A_+$ vanishes and thus $f_{++}^*=f_{-+}^*$. Consequently at that point ${\rm FP}_1$ and ${\rm FP}_2$ merge. Increasing $\xs$ further produces an imaginary $A_+$ and these FPs migrate to the complex plane. Similarly, at the value
\begin{equation}
\label{crit2}
\xs^{(2)}\simeq0.8403
\end{equation}
${\rm FP}_3$ and ${\rm FP}_4$ collide due to the vanishing of $A_-$ and beyond that point they become complex. This means that, for ratios above this value, the model has four cFPs in two conjugate pairs. Their location can be seen in Fig.~\ref{fig:lhf}.

Interestingly, these pairs of cFPs come close to one another if we continue to increase 
$\xs$ and they coincide at the critical value $\xc=2$, where $B=0$ and $A_+=A_-$\footnote{This exact coincidence of the cFPs is achieved only at large $\nc$. At finite $\nc$ the distance between their locations is $O(1/\nc)$ for $\xs=2$.}. At $\xs=\xc$ ${\rm FP}_1$ collides with  ${\rm FP}_3$ and ${\rm FP}_2$ collides with ${\rm FP}_4$. We can think of $|\xs-\xc|$ roughly as the analog of the parameter $\delta$ in the previous section. The analysis in \cite{Hansen:2017pwe} seems to show that for such high values of $\xs$ the model is no longer asymptotically free and would need to be UV completed. However, if the completion happens at sufficiently high energies the physics of interest here should remain qualitatively unaffected. 

Leaving aside the 't Hooft coupling, that we set to its FP value $\lambda^*$, the conformal dimensions of the other two operators at the different FPs are
\begin{align}\label{eq:dimBIS}
\Delta_{{h}}\left({\rm FP}_1,{\rm FP}_2\right)&=4+8\lambda^* B\,,\qquad\qquad \qquad\Delta_{\tilde{f}}\left({\rm FP}_1,{\rm FP}_2\right)=4\pm8\lambda^*A_+\,,\nonumber\\[3mm]
\Delta_{{h}}\left({\rm FP}_3,{\rm FP}_4\right)&=4-8\lambda^* B\,,\qquad\qquad \qquad\Delta_{\tilde{f}}\left({\rm FP}_3,{\rm FP}_4\right)=4\pm8\lambda^*A_-\,.
\end{align}
Since the $\beta$-functions $\beta_h, \beta_f$ are not diagonal in the couplings $h,f$, the operators that diagonalize the dilatations at these FPs are not directly those in \eqref{operators}. However, the fact that $f$ does not appear in $\beta_h$ means that we can simply define a new coupling $\tilde{f}=f + \alpha h$, with $\alpha$ an appropriate constant, in such a way that the beta functions $\beta_h, \beta_{\tilde{f}}$ become diagonal in $h, \tilde{f}$ at the FPs. At the level of operators this means that, in terms of operators with well defined dimensions \eqref{eq:dimBIS}, the original combination becomes
\be
h \mathcal{O}_h + f \mathcal{O}_f = h \tilde{\mathcal{O}}_h + \tilde{f} \mathcal{O}_f \,,\qquad 
\tilde{\mathcal{O}}_h = \mathcal{O}_h - \alpha \mathcal{O}_f \,.
\ee 
In other words, the dimension $\Delta_h$ in \eqref{eq:dimBIS} is not the dimension of $\mathcal{O}_h$ but that of the linear combination $\tilde{\mathcal{O}}_h$. Nevertheless, the running of the corresponding coupling is still given by $\beta_h$. In contrast, the dimension 
$\Delta_{\tilde{f}}$ corresponds to that of the original double-trace operator $\mathcal{O}_f$, but the corresponding coupling $\tilde{f}$ and its $\beta$-function are linear combinations of $h,f$ and $\beta_h, \beta_f$, respectively. 

 The dimensions \eqref{eq:dimBIS} depend on $\lambda^*$ and $\xs$ but not on $\xf$. This means that, in the following, we can assume that we adjust $\xf$ so that $\lambda^*$ is small and our  perturbative treatment is reliable, as anticipated below \eqref{lambda}. Moreover, the dependence on $\lambda^*$ in \eqref{eq:dimBIS} is extremely simple: $\lambda^*$ simply controls the overall size of the imaginary part of the dimensions, so it plays the role of the parameter $\epsilon$ in  previous sections. The non-trivial dependence on $\xs$ enters through the functions $A_\pm(\xs)$ and $B(\xs)$ given in \eqref{AB}. This dependence  implies that $\tilde{\mathcal{O}}_f$ becomes marginal when two real FPs merge, whereas 
 $\mathcal{O}_{{h}}$ becomes marginal when two pairs of cFPs merge. The first property follows from the fact that the real mergers ${\rm FP}_1 \leftrightarrow {\rm FP}_2$  and ${\rm FP}_3 \leftrightarrow {\rm FP}_4$ happen at $\xs=\xs^{(1)}$  and $\xs=\xs^{(2)}$, respectively. Since $A_+(\xs^{(1)})=A_-(\xs^{(2)})=0$ we conclude that 
 $\Delta_f=4$ at both real mergers. In contrast, at the merger of the two pairs of cFPs,  
 ${\rm FP}_1 \leftrightarrow {\rm FP}_3$  and ${\rm FP}_2 \leftrightarrow {\rm FP}_4$, 
  we have $\xs=\xc$ and $B(\xc)=0$, so at this point $\Delta_{{h}}=4$. According to our general analysis, we expect a change of scaling at this point.
  
In order to see this it is useful to rewrite the $\beta$-functions in terms of the difference $\xs-\xc$:
\begin{equation}
\begin{aligned}
\beta_h&=4\left[3+\left(\xs-\xc\right)\right]h^2-6\,\lambda^*\, h+\frac34(\lambda^*)^2\,,\\[2mm]
\beta_f&=4f^2+8\left[3+\left(\xs-\xc\right)\right]f h+12 \left[2+\left(\xs-\xc\right)\right] h^2-6\lambda^* f+\frac{3}{4}\left[2+\left(\xs-\xc\right)\right](\lambda^*)^2\,.
\end{aligned}
\end{equation}
The $\beta$-function for the coupling $h$  decouples from that  for $f$, so it can be readily integrated. The running of the coupling with the energy scale $\mu$, in the regime $\xs>\xc$ we are interested in, is thus
\begin{equation}\label{eq:runh}
h=\lambda^*\,\frac{3+\sqrt{3\left(\xs-\xc\right)}\,\tan\left(\lambda^*\sqrt{3\left(\xs-\xc\right)}\,\log\frac{\mu}{\mu_0}\right)}{12+4\left(\xs-\xc\right)}\,,
\end{equation}
for some reference scale $\mu_0$. This relation can be inverted to obtain the energy scale in terms of the coupling. Alternatively, it is possible to directly integrate the (inverse of) $\beta_h$ as in previous sections to get the following relation between the UV and IR scales, 
$\mu_{\text{\tiny UV}}$ and $\mu_{\text{\tiny IR}}$, and their associated couplings, $h_{\text{\tiny UV}}$ and $h_{\text{\tiny IR}}$:
\begin{equation}
\begin{aligned}
\log \frac{\mu_{\text{\tiny UV}}}{\mu_{\text{\tiny IR}}}=\frac{1}{\lambda^*\sqrt{3\left(\xs-\xc\right)}}&\left[\arctan\left(\frac{4\left[3+\left(\xs-\xc\right)\right]\left(h_{\text{\tiny UV}}-h_{\text{\tiny max}}\right)}{\lambda^*\sqrt{3\left(\xs-\xc\right)}}\right)\right.\\[2mm]
&\left.-\arctan\left(\frac{4\left[3+\left(\xs-\xc\right)\right]\left(h_{\text{\tiny IR}}-h_{\text{\tiny max}}\right)}{\lambda^*\sqrt{3\left(\xs-\xc\right)}}\right)\right]\,,
\end{aligned}
\end{equation}
where 
\begin{equation}
h_{\text{\tiny max}}=\frac{3\lambda^*}{4\left[3+\left(\xs-\xc\right)\right]}
\end{equation}
is the position of the maximum of the $\beta$-function. In the approximation 
\be
\left|h_{\text{\tiny UV,IR}}-h_{\text{\tiny max}}\right|\gg\lambda^*\sqrt{\left(\xs-\xc\right)}
\ee
this reduces to 
\begin{equation}
\label{therefore}
\left. \log \frac{\mu_{\text{\tiny UV}}}{\mu_{\text{\tiny IR}}} \right|_{h}\simeq\frac{\pi}{\lambda^*\sqrt{3\left(\xs-\xc\right)}}\,.
\end{equation}
At leading order in the limit $\xs-\xc\ll1$ the dimension $\Delta_{{h}}$ in Eq. \eqref{eq:dimBIS} is 
\be
\Delta_{{h}}\simeq 4\pm i\,2\,\lambda^*\sqrt{3\left(\xs-\xc\right)} \,. 
\ee
We can therefore rewrite \eqref{therefore} as
\begin{equation}
\left. \log \frac{\mu_{\text{\tiny UV}}}{\mu_{\text{\tiny IR}}} \right|_{h} \simeq 
\frac{2\pi}{|\operatorname{Im}{\Delta}_{{h}}|} \,.
\end{equation}
This result is again the Miransky scaling  \eqref{eq:Miransky}, but in this case we have two independent small parameters controlling the hierarchy: $\lambda^*$ and $\xs-\xc$ or, equivalently, $\nf$ and $\ns$. As we have mentioned, these are akin to $\epsilon$ and $\delta$ in  \Sec{sec:toy}. The extra factor of $\sqrt{\xs-\xc}$ in the denominator of  \eqref{therefore} is due to the fact that ${h}$ is marginal at $\xc$.

The $\beta$-function for $f$ is  more complicated, since it involves also the coupling $h$ that we have just studied. Fortunately, for our purposes it is enough to fix this coupling to its conformal value at $\xs=\xc$. Expanding the complete result \eqref{eq:runh} it can be seen that 
\begin{equation}
h=\frac{\lambda^*}{4}+O\left(\xs-\xc\right)\,.
\end{equation}
The $O\left(\xs-\xc\right)$ is a small correction in the region of interest here in which $h$ is walking. Including this correction  only gives a subleading contribution to the scaling of $f$. Indeed, substituting this in $\beta_f$ and integrating we find 
\begin{equation}
f\simeq\lambda^*\frac{\sqrt{3}}{2}\tan\left(2\sqrt{3}\,\lambda^*\,\log\frac{\mu}{\mu_0}\right)+O\left(\xs-\xc\right)\,.
\end{equation} 
Following the same steps as before we obtain the hierarchy 
\begin{equation}
\left. \log \frac{\mu_{\text{\tiny UV}}}{\mu_{\text{\tiny IR}}} \right|_{f}\simeq\frac{\pi}{2\sqrt{3}\lambda^*}\,.
\end{equation}
Taking into account that the dimension of the double-trace operator at the fixed points is 
\begin{equation}
\Delta_{\tilde{f}}=4\pm i\,4\sqrt{3}\,\lambda^*+O\left(\lambda^*\sqrt{\xs-\xc}\right)
\end{equation}
we can rewrite this result as
\begin{equation}
\label{bandana}
\left. \log \frac{\mu_{\text{\tiny UV}}}{\mu_{\text{\tiny IR}}} \right|_{f} \simeq 
\frac{2\pi}{|\operatorname{Im}{\Delta}_{\tilde{f}}|} \,,
\end{equation}
recovering the expected Miransky scaling. 

\begin{figure}[t]
\begin{center}
		\begin{subfigure}{0.46\textwidth}
			\includegraphics[width=\textwidth]{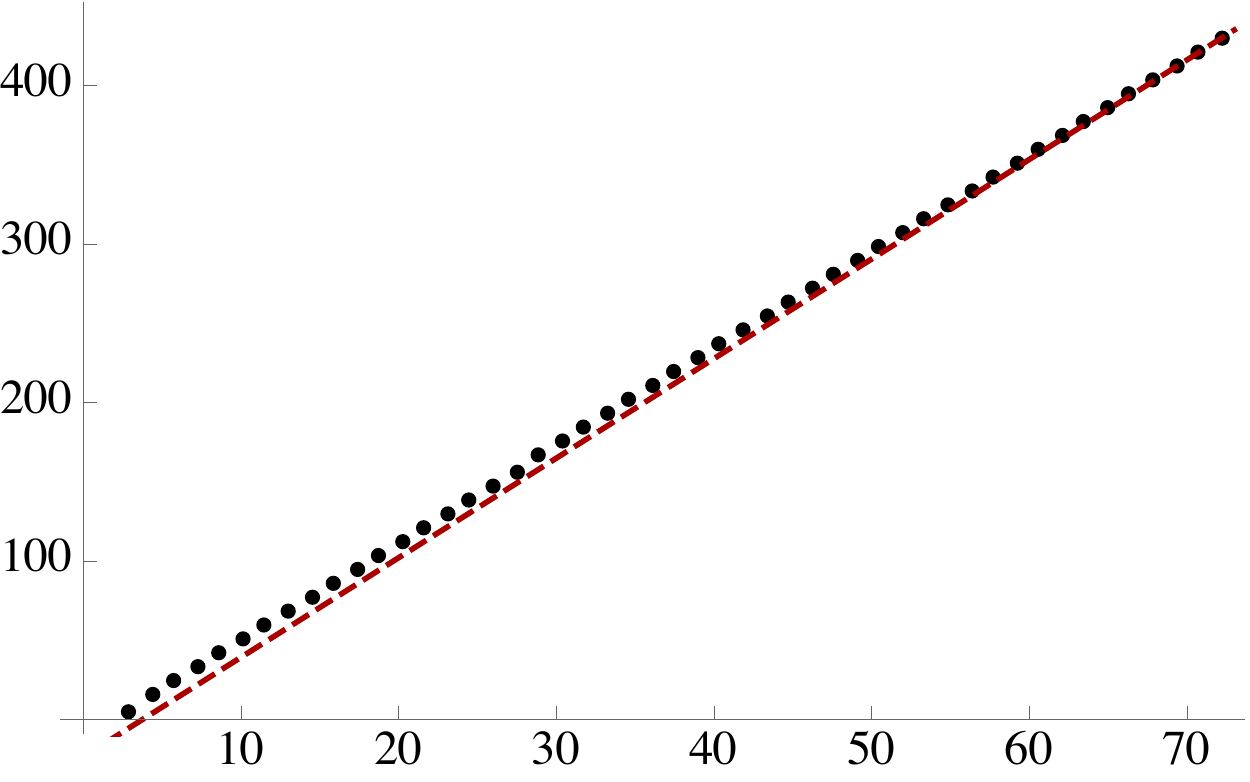} 
			\put(-185,115){$\left. \log \frac{\mu_{\text{\tiny UV}}}{\mu_{\text{\tiny IR}}} \right|_{f}$}
			\put(-35,18){$|{\rm Im}\Delta_{\tilde{f}}|^{-1}$}
		\end{subfigure}\hspace{5ex}
		\begin{subfigure}{.46\textwidth}
			\includegraphics[width=\textwidth]{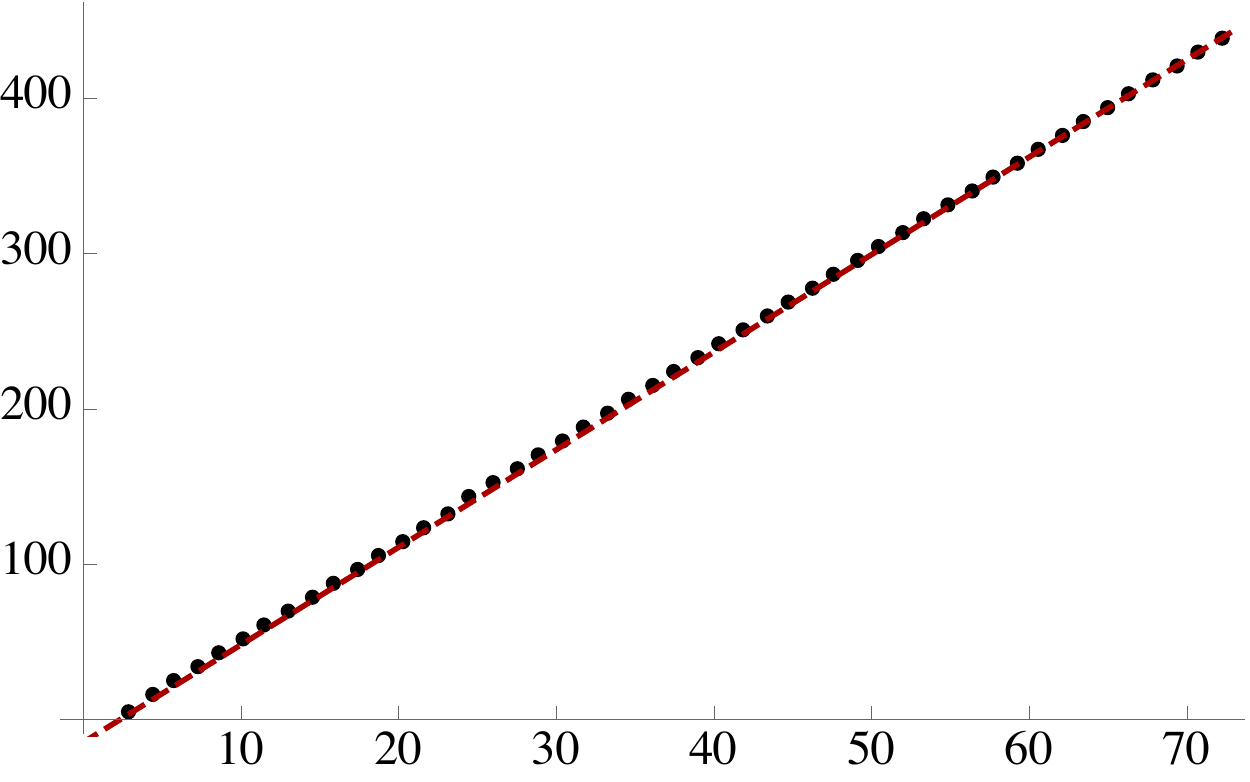} 
				\put(-185,115){$\left. \log \frac{\mu_{\text{\tiny UV}}}{\mu_{\text{\tiny IR}}} \right|_{f}$}
			\put(-35,18){$|{\rm Im}\Delta_{\tilde{f}}|^{-1}$}
		\end{subfigure}
		\caption{\small Hierarchy of scales as a function of $|{\rm Im}\,\Delta_{\tilde{f}}|^{-1}$ for $\xs-\xc=0.15$ (left) and $\xs-\xc=0.05$ (right). The dashed red line has slope $2\pi$ and passes through the last point computed numerically. The scaling \eqref{bandana} is independent of $\xs$ but it is satisfied at smaller values of $\lambda^*$ for larger $\xs-\xc$.}
		\label{fig:scalingBIS}
	\end{center}
\end{figure}

As a check of these results we have also solved numerically the $\beta$-function for $f$ using the complete running of $h$ as given in $\eqref{eq:runh}$. Following a similar strategy to that in \Sec{sec:toy}, we declare that the coupling is ``walking'' whenever the $\beta$-function is smaller than some control parameter, which was fixed to $10^{-2}$ in our example. This determines the UV and IR scales between which the scaling should be satisfied. Repeating the integration for several values of $\lambda^*\ll1$ for fixed $\xs-\xc\ll1$ one obtains curves as those in Fig. \ref{fig:scalingBIS}, where we see that the scaling \eqref{bandana} is verified for small $\lambda^*$, corresponding to large $|{\rm Im}\,\Delta_{\tilde{f}}|^{-1}$, independently of $\xs-\xc$.  

The conclusion of the analysis above is that the hierarchy associated to the walking of the coupling $h$ is enhanced by a factor $1/\sqrt{\left(\xs-\xc\right)}$ with respect to that of the coupling $f$. This means that, assuming that both couplings start walking at the same UV scale, the double trace operator stops walking at an energy scale $\mu_f$ that is parametrically higher than the scale $\mu_h$ at which $h$ stops walking. Specifically, when $\xs$ is close to $\xc$ the ratio between these hierarchies is 
\begin{equation}
\log \frac{\mu_{{\tiny f}}}{\mu_{{ {h}}}} \simeq \frac{\pi}{\sqrt{3\left(\xs-\xc\right)}\,\,\lambda^*}\,.
\end{equation}
The fact that $h$ walks much more slowly than $f$ is the intuitive reason why the walking region of $f$ can be obtained by setting $h$ to its FP value.

\subsubsection{A Landau--Ginzburg--Wilson example}

We now turn to a second example in a theory with ${\rm O}(n)\times{\rm O}(m)$ symmetry. These type of models, dubbed Landau--Ginzburg--Wilson, are relevant in spin systems, where both semi-simple group factors are global. It is possible to make one of the factors local by introducing a vector field through the usual gauging procedure, so we consider a set of $m$ vectors of dimension $n$, $\vec{\phi}_\alpha=\left\{\phi_{\alpha a}\right\}$ with $\alpha=1,\dots,m$ and $a=1,\dots,n$, together with a gauge field $A^{\alpha\beta}_\mu$, antisymmetric in $\alpha$ and $\beta$, interacting in $d=4-\epsilon$ dimensions via the Euclidean Lagrangian \cite{Pelissetto:2001fi,Bonati:2020kks}
\begin{equation}\label{eq:Lagrangian}
\mathcal{L}=\frac14F_{\mu\nu}^2+\frac12\left(\partial_\mu\vec{\phi}_\alpha+gA_\mu^{\alpha\beta}\vec{\phi}_\beta\right)^2+\frac{1}{24}u\left(\vec{\phi}_\alpha^{\,2}\right)^2+\frac{1}{24}v\left[\left(\vec{\phi}_\alpha\cdot\vec{\phi}_\beta\right)^2-\vec{\phi}_\alpha^{\,2}\vec{\phi}_\beta^{\,2}\right]\,.
\end{equation}
We have omitted a mass term for the scalars but we have allowed for two different quartic couplings compatible with the symmetry. The system enjoys local ${\rm O}(m)$ invariance supplemented with a global ${\rm O}(n)$ symmetry. Models in this class may have phenomenological relevance in solid state physics, as suggested for instance in \cite{Sachdev:2018nbk}.

We will continue the model to non-integer values of $n$ and $m$. Although we are not in the large-$N$ limit, this is commonly considered in the condensed matter literature in the same way as non-integer dimensionalities. See \cite{Binder:2019zqc} for a proposal to make the analytic extension to non-integers rigorous.\footnote{Although their proposal is for global, not gauged, symmetries.}

At one loop, the system of $\beta$-functions for the gauge coupling $\alpha=g^2$ and quartic couplings reads \cite{Pelissetto:2001fi,Bonati:2020kks}
\begin{align}
\beta_\alpha&=-\epsilon\,\alpha+\left[\frac{n}{12}-\frac{11}{3}\left(m-2\right)\right]\alpha^2\,,\nonumber\\[2mm]
\beta_v&=-\epsilon\,v+\frac{m+n-8}{6}v^2+2u\,v-\frac32\left(m-1\right)v\,\alpha+\frac94\left(m-2\right)\alpha^2\,,\\[2mm]
\beta_u&=-\epsilon\,u+\frac{m\,n+8}{6}u^2+\frac{\left(m-1\right)\left(n-1\right)}{12}\left(v^2-2u\,v\right)-\frac32\left(m-1\right)u\,\alpha+\frac98\left(m-1\right)\alpha^2\,.\nonumber
\end{align}
As usual, the $\beta$-function for the gauge coupling decouples at this level. It gives, apart from a Gaussian fixed point, a Wilson--Fisher-type point at
\begin{equation}
\alpha^*=\frac{12}{88-44m+n}\,\epsilon\,,
\end{equation} 
which can be made arbitrarily close to the Gaussian point by tuning the dimensionality of the space. We will be interested in the set of fixed points resulting from substituting this into the $\beta$-functions for $u$ and $v$. First, notice that for $n=1$ the equation for $u$ decouples. This limit is expected to be somewhat singular, since for ${\rm O}(1)=\mathset{Z}_2$ global symmetry there should be a single quartic coupling. In other words, the term in square brackets accompanying $v$ in \eqref{eq:Lagrangian} vanishes identically. However we do not detect any problem at the level of the $\beta$-functions. Although the theory is not well defined at exactly $n=1$, we will circumvent this issue by taking $n$ close to one and ignoring corrections depending on $n-1\ll 1$.

\begin{figure}[t]
\begin{center}
		\begin{subfigure}{0.44\textwidth}
			\includegraphics[width=\textwidth]{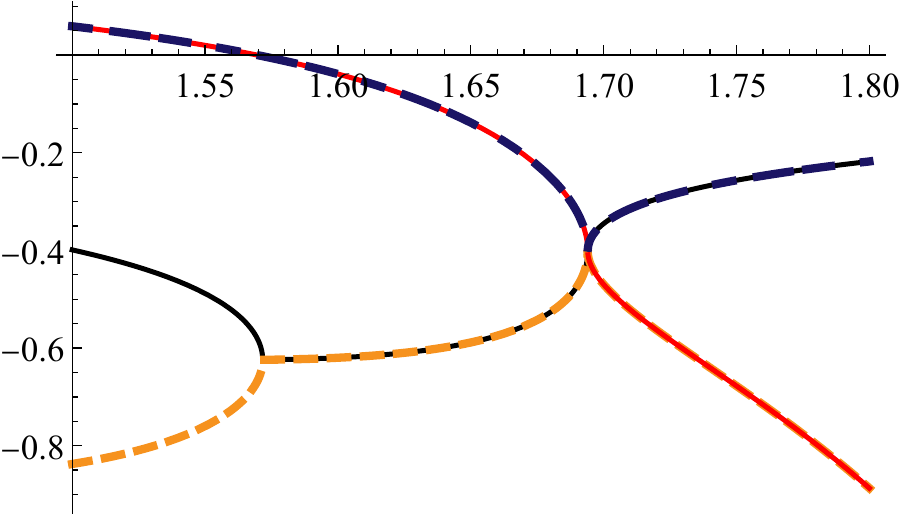} 
			\put(-190,120){Re\,$v$}
			\put(3,95){$m$}
		\end{subfigure}\hspace{10ex}
		\begin{subfigure}{.44\textwidth}
			\includegraphics[width=\textwidth]{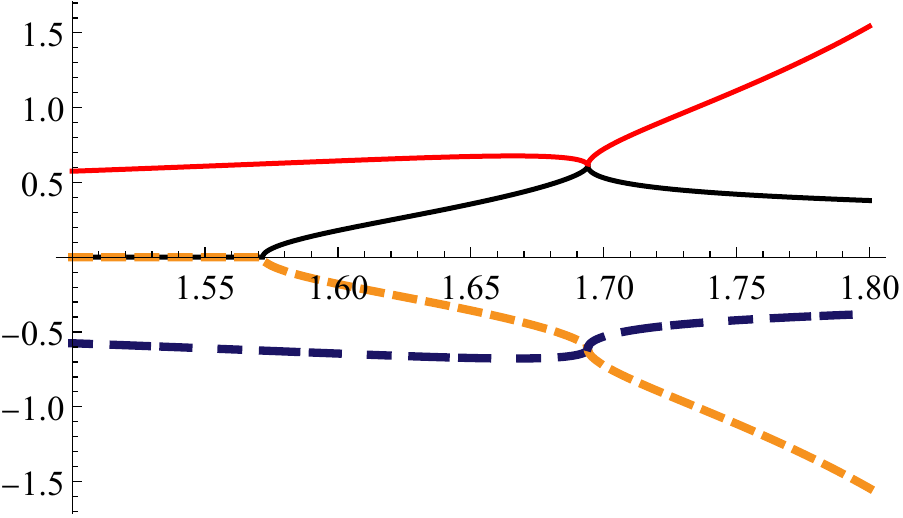} 
			\put(-190,120){Im\,$v$}
			\put(3,50){$m$}
		\end{subfigure}
		\caption{\small The real (left) and imaginary (right) parts of the coupling $v$ at the four fixed points as a function of $m$ for $n=1$.}
		\label{fig:vFP}
	\end{center}
\end{figure}

We thus find the fixed points
\begin{equation}
u^*_{\pm}=\frac{3\left(26m-71\pm\sqrt{568m^2-4448m+5905}\right)}{(m+8)(44m-89)}\,\epsilon\,.
\end{equation}
For each of these solutions, there are two zeroes in the $\beta$-function for $v$, so there is a total of four fixed points, all of them proportional to $\epsilon$ and therefore perturbative close to four dimensions. We can study their position in terms of $m$. They are all complex (in two complex conjugate pairs) for $m>-0.857\dots$, so no real points exist for positive $m$. There is a critical value
\begin{equation}
m_{\text{\tiny{c}}}=\frac{1112-27\sqrt{546}}{284}\approx1.694
\end{equation}
at which $u^*_+=u^*_-$ and the associated operator becomes marginal. For $m>m_{\text{\tiny{c}}}$ the coupling $u$ becomes complex. Interestingly, at the critical value the two pairs of complex-conjugate fixed points coincide, as can be seen in Fig.~\ref{fig:vFP} for the coupling $v$.  Note that for negative values of $v$ there may be an instability, depending on the value of $u$.

Picking a larger value of $n$ separates these pairs of fixed points, but their mutual influence can still be felt if it is close enough to $n=1$. As expected, there is a hierarchy of hierarchies as in the previous example. While for the coupling $v$ the size of the walking region grows exponentially as $\sim 1/\epsilon$, this is increased to $\sim  1/(\epsilon\,\sqrt{m-m_{\text{\tiny{c}}}})$ for $u$.

\section{Holographic realization}
\label{sec:strong}

A holographic realization of the mechanism of fixed point annihilation and complex conformal field theories was proposed in \cite{Faedo:2019nxw}. The gravitational theory must contain a parameter, playing the role of $\epsilon$ in \eqref{eq:betasimp}, such that by tuning it we can change the extrema of the potential from real to complex values of the scalars. Extrema of the potential corresponding to a negative cosmological constant give rise to AdS$_{d+1}$ solutions, whose field theory duals are $d$-dimensional CFTs. If the extrema are located at complex values of the scalars, the associated AdS geometry would generically have a complex radius, resulting in complex conformal dimensions for the operators, complex central charge,  etc. The field theory duals to these solutions would be the complex conformal field theories of \cite{Gorbenko:2018ncu, Gorbenko:2018dtm}. 

In order to have several pairs of complex fixed points and study their interplay one needs a potential admitting various extrema, as well as enough parameters to move around their location on the complex plane. We will make two simplifying assumptions. First, we will assume that a single operator is driving the RG flow, as in \Sec{sec:toy}, and therefore consider a unique scalar in the $(d+1)$-dimensional gravity dual with action
\begin{equation}\label{eq:action}
S\,=\,\frac{1}{2\kappa^2}\int d^{d+1}x\sqrt{-g}\left[R-\frac12 (\partial\phi)^2-V(\phi)\right]\,.
\end{equation}
Second, we will work at the level of a superpotential from which the potential is deduced (see Eq.~\eqref{eq:potential} from Appendix~\ref{sec:arbitrary}). This is not necessary but it is  technically  convenient.\footnote{In particular, this implies that we only consider solutions to the first-order equations \eqref{eq:BPS}, which are a subset of complete set of solutions to the second-order Einstein's equations coming from \eqref{eq:action}.}

We will study in detail the simplest case of two pairs of complex-conjugate fixed points. Since extrema of the superpotential are also extrema of the potential, let us consider a superpotential $W$ whose derivative is given by
\begin{equation}\label{eq:derivativeW}
W'(\phi) = \frac{W_0}{L}\ \phi \,(\phi -\phi_1) \ (\phi -\bar \phi_1) \ (\phi -\phi_2)\ (\phi -\bar \phi_2)\,.
\end{equation}
There is a (real) fixed point at $\phi=0$ that we included in order to provide a UV completion. As we will see, this has consequences in the resulting scaling. The other fixed points, located at $\phi=\{\phi_n, \bar{\phi}_n\}$ with $n=1,2$, can be made real or complex depending on the value of the appropriate parameters. Notice that the superpotential itself must be real, so if any of the $\phi_n$ is complex then necessarily $\bar{\phi}_n=\phi_n^*$. The constant $W_0$ together with the constant of integration to obtain $W$ from \eqref{eq:derivativeW} will be fixed so that the operator has conformal dimension $\Delta_{\text{\tiny UV}}<d$ at the UV fixed point $\phi=0$, with $L$ the radius of its dual AdS solution. The $\beta$-function associated to this superpotential can be computed as 
\begin{equation}
\beta\left(\phi\right)=-2\left(d-1\right)\frac{W'(\phi)}{W(\phi)}=-2\left(d-1\right)\frac{\dd \log{W}}{\dd\phi}\,.
\end{equation}
It vanishes at the same points where the derivative of the superpotential vanishes, corresponding to AdS solutions and therefore to dual conformal field theories. More details can be found in Appendix \ref{sec:ED}.

The way in which a model of this kind captures the physics of fixed point annihilation is as follows. Let us focus on the first pair of extrema at $\phi_1$ and $\bar{\phi}_1$. When both of these are real they correspond to a pair of real fixed points. If they are close to each other, in their vicinity the $\beta$-function looks like the top curve in the left panel of Fig.~\ref{fig:annihilation}. Suppose now that by tuning a parameter in the superpotential we can achieve $\phi_1=\bar{\phi}_1$. The fixed points merge, the extremum is then double, the dual operator is marginal and we are in the middle curve of the same figure. Finally, if varying again the parameter the extrema can be made complex conjugate to each other, $\bar{\phi}_1=\phi_1^*$, the fixed points have migrated into the complex plane and the lower curve is reproduced. Clearly the same process can be repeated with the second pair, $\phi_2$ and $\bar{\phi}_2$, so the complete superpotential \eqref{eq:derivativeW} is suited to capture the right panel of Fig.~\ref{fig:annihilation} and the type of physics we have considered in \Sec{sec:toy}.

To see this we parametrize the four fixed points as 
\begin{equation}
\phi_{1,2}= \phi_0\pm \delta + i\epsilon\qquad\qquad\text{and}\qquad\qquad\bar{\phi}_{1,2}=\phi_{1,2}^*=\phi_0\pm \delta - i\epsilon\,.
\end{equation}
We assume $|\phi_0|\gg \delta, \epsilon$. When $\epsilon=0$ and $\delta\ne0$ we have two real double fixed points. These fixed points merge at $\phi_0$ if we adjust $\delta=0$, and migrate into the complex plane as two double cFPs when $\delta=0$ but $\epsilon\ne0$, reproducing the three situations in the right panel of Fig.~\ref{fig:annihilation}. Alternatively, we can fix $\epsilon$ non-vanishing but small and study how the system evolves as $\delta$ decreases and the cFPs 
approach each other along a direction parallel to the real axis, as in \Sec{sec:toy}.

A crucial difference between this holographic model and the one discussed in \Sec{sec:toy} is the presence of a real UV fixed point besides the complex ones. This has no effect on the RG flow when the two pairs of cFPs are well separated, that is, if $\delta\gg\epsilon$. In this case the scaling is given by \eqref{eq:twice}, as in $\textit{\textbf{Case A}}$ of the simple model of \Sec{sec:toy}.

\begin{figure}[h!!!]
	\begin{center}
		\begin{subfigure}{.48\textwidth}
			\includegraphics[width=1\textwidth]{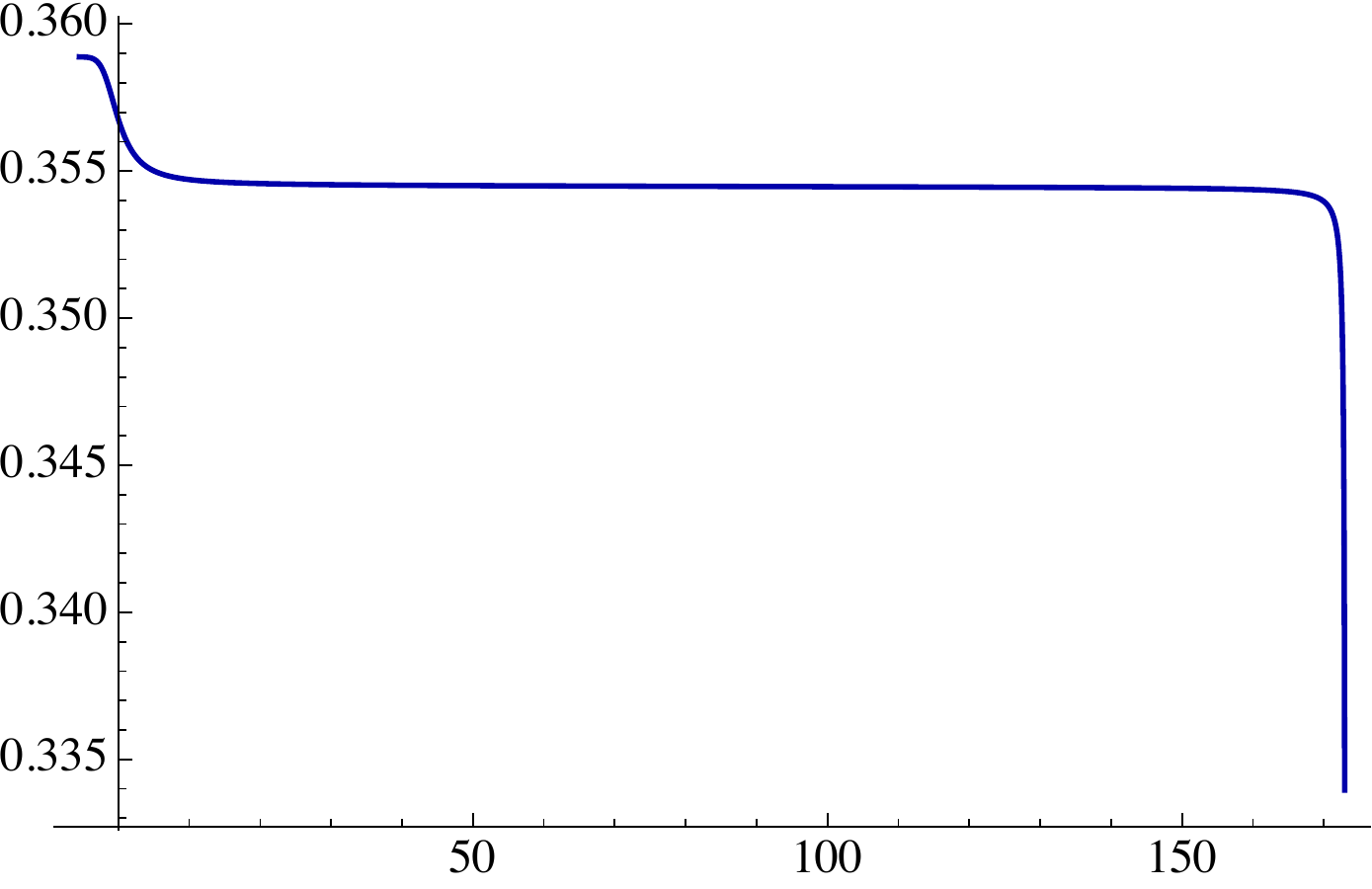}
			\put(-20, -10){$ \log \ell $}
			\put(-185, 130){$ \log (F_{q\bar q}\, \ell^2) $}
		\end{subfigure}
		\hfill
		\begin{subfigure}{.48\textwidth}
			\includegraphics[width=1\textwidth]{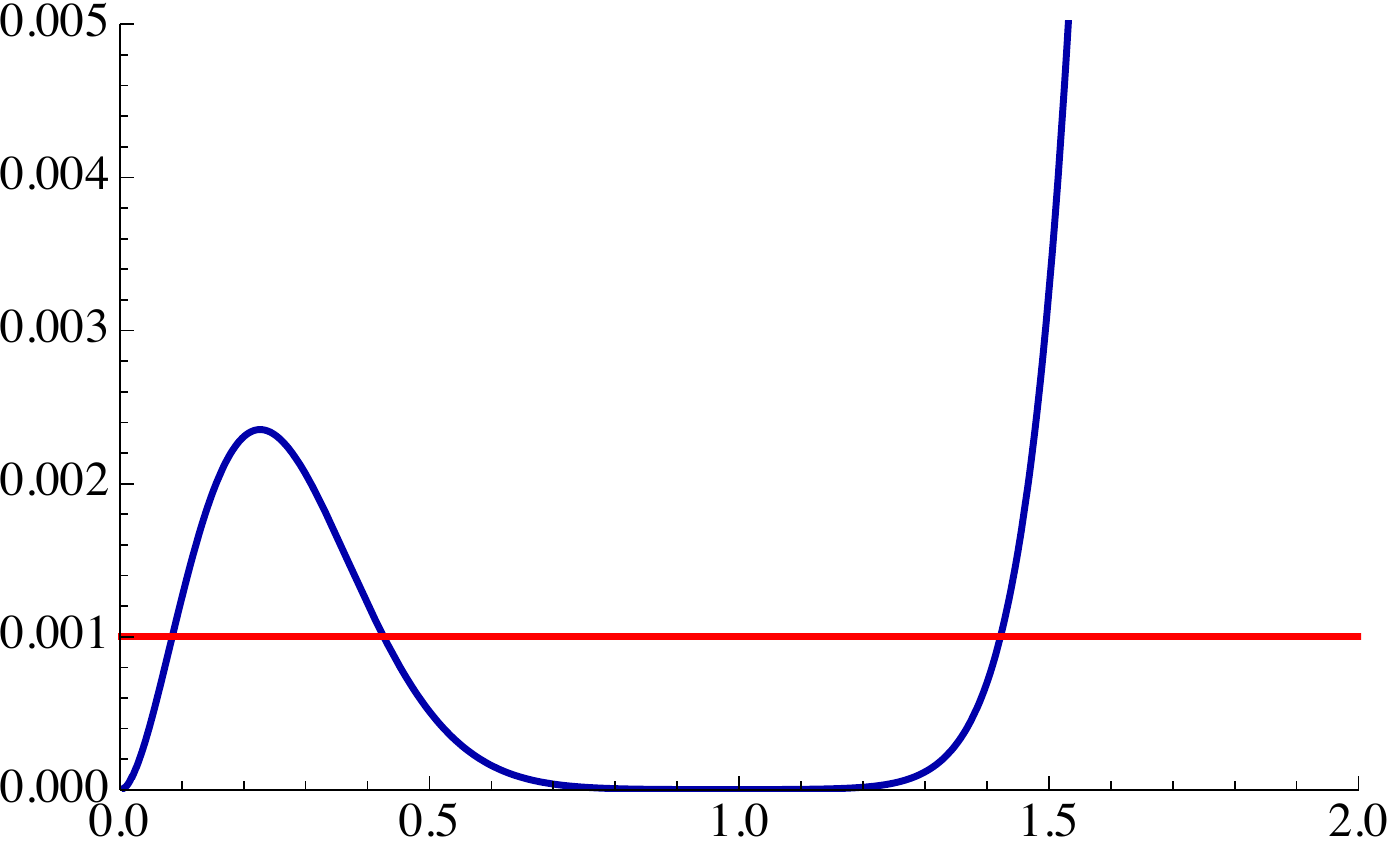}
			\put(-10, -10){$ \phi_* $}
			\put(-190, 125){$\left| \frac{\dd \log (F_{q\bar q}\, \ell^2 )}{\dd \log \ell} \right|$}
		\end{subfigure}
		\begin{subfigure}{.48\textwidth}
			\includegraphics[width=1\textwidth]{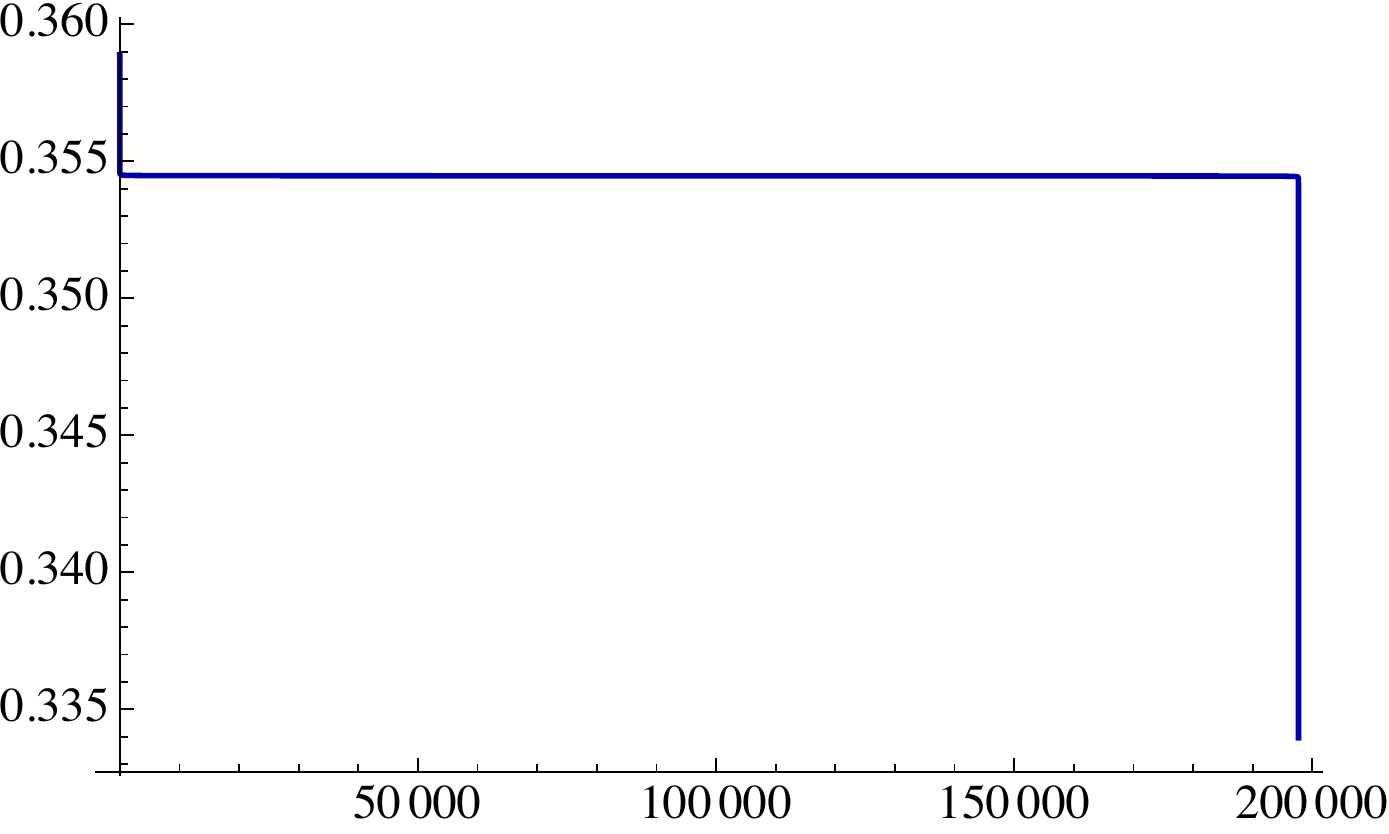}
			\put(-20, -10){$ \log \ell $}
			\put(-185, 130){$ \log (F_{q\bar q}\, \ell^2  ) $}
		\end{subfigure}
		\hfill
		\begin{subfigure}{.48\textwidth}
			\includegraphics[width=1\textwidth]{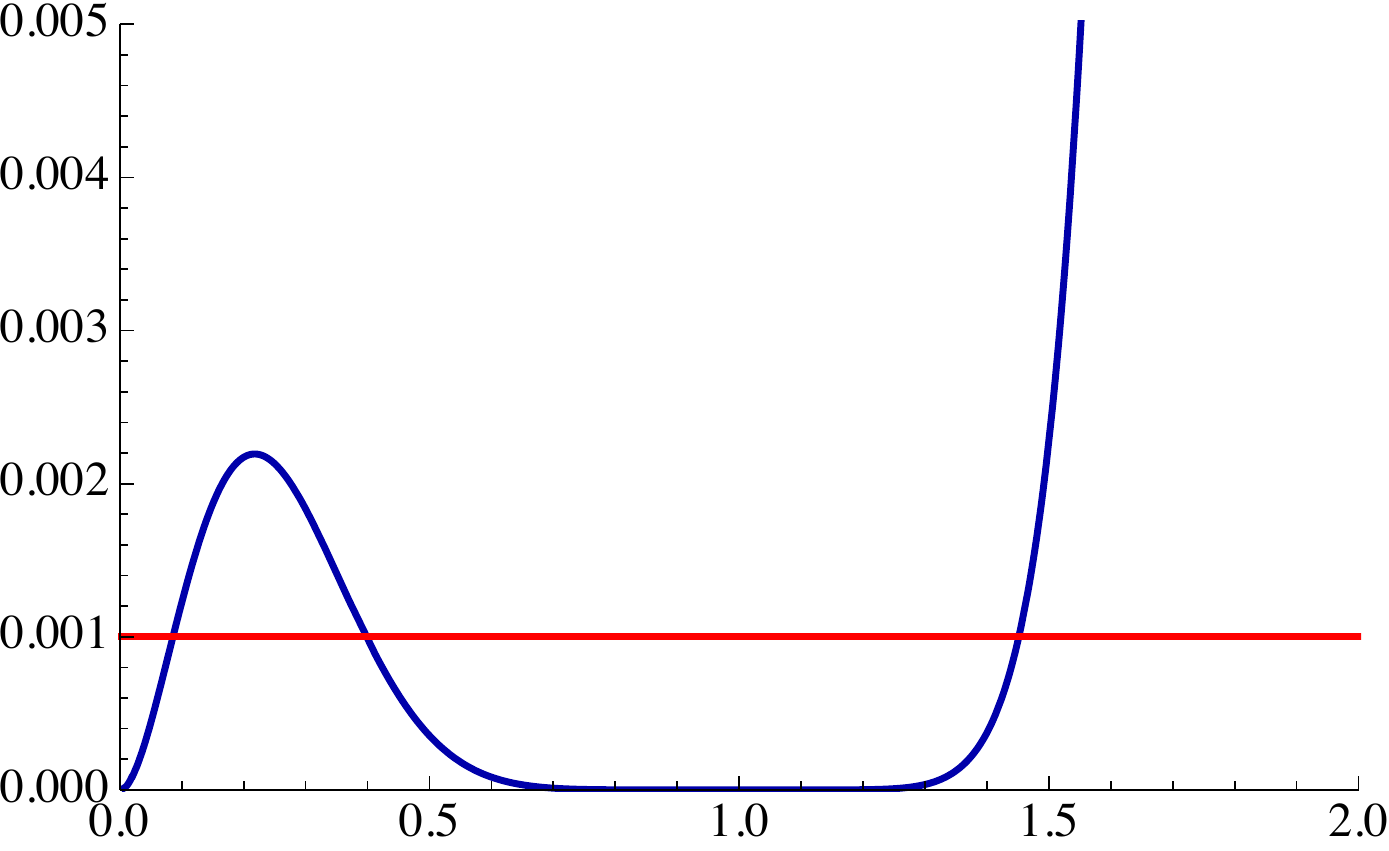}
			\put(-10, -10){$ \phi_*$}
			\put(-190, 125){$ \left|\frac{\dd \log (F_{q\bar q}\, \ell^2)}{\dd \log \ell} \right|$}
		\end{subfigure}
		\caption{\small Quark-antiquark force $F_{q\bar q}\ell^2$ as a function of the separation $\ell$ (left) and the derivative defined in \eqref{eq:thereshold} as a function of the turning point $\phi_*$ defined in Appendix \ref{sec:ED} (right). The two different cases are $\{\epsilon,\delta\} = \{0.2,\ 0.1\}$ (top) and $\{\epsilon,\delta\}= \{0.02,\ 10^{-7}\}$ (bottom). The red line stands for the tolerance $\nu= 10^{-3}$ below which we encounter the walking region. This gives $\phi_{\text{\tiny UV}}$ and $\phi_{\text{\tiny IR}}$ from which we get $\ell_{\text{\tiny UV}}$ and $\ell_{\text{\tiny IR}}$ using \eqref{eq:Lfromphi}. The region below the red line close to the origin corresponds to the UV fixed point.}
		\label{fig:walkingregime}
	\end{center}
\end{figure}
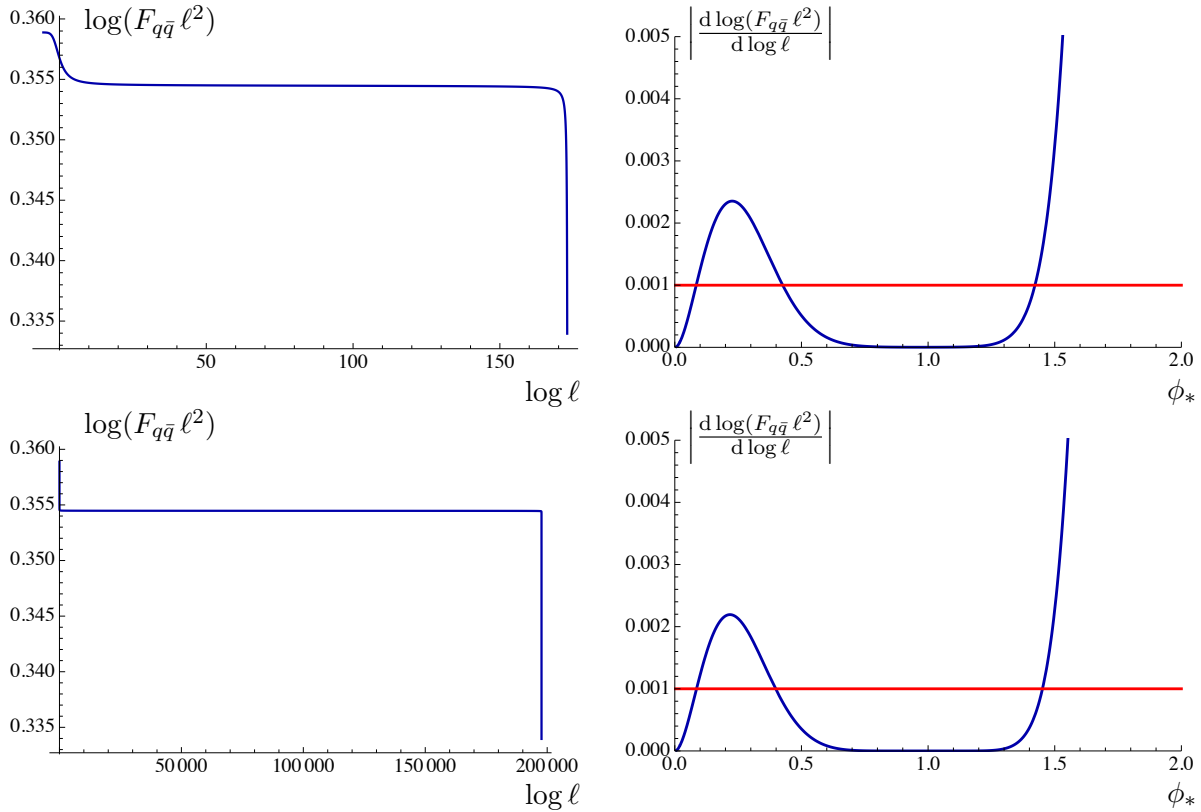

However, in the opposite limit, $\delta\ll\epsilon < 1$, it is convenient to  consider the imaginary part of the dimension of the operator at both cFPs, located at $\phi_0 \pm \delta + i\epsilon$, which we denote Im$\,\Delta_\pm$. As explained in Appendix \ref{sec:arbitrary}, this leads to the hierarchy 
\begin{equation}\label{eq:newscaling}
\log \, \frac{\mu_{\text{\tiny UV}}}{\mu_{\text{\tiny IR}}}\ \approx \ \frac{2\pi}{\text{Re}(d-\Delta_\pm)^2}\, \Big|\, \text{Im}\,\Delta_- \, +\,  \text{Im}\,\Delta_+\, \Big|
\end{equation}
Note that this scaling reduces to \eqref{eq:scalingC} when Im$\,\Delta_+ = \text{Im}\,\Delta_-$ at leading order, as in the model considered in that Section.

Thus, in this particular case and in the presence of a (real) UV fixed point, the hierarchy when $\delta\ll\epsilon$ is again $O\left(1/\epsilon^3\right)$ but the expression is a little bit more involved and depends on data at both cCFTs. In other words, the UV fixed point induces a small distinction between $\Delta_-$ and $\Delta_+$, the sum of the imaginary part of which is now determining the scaling.

We now check these results using holography. First of all, we need a way to characterize the size of the walking region, that is, the energy scales $\mu_{\text{\tiny UV}}$ and $\mu_{\text{\tiny IR}}$ between which some form of scaling is satisfied. In this energy interval the theory is quasi-conformal, so a natural possibility is to analyze whether a certain physical quantity behaves as in a CFT. In \cite{Faedo:2019nxw} we used the temperature dependence of the entropy density, which verifies $S\propto T^{d-1}$ in a $d$-dimensional CFT. To give a complementary check of the holographic approach, we will analyze here the Wilson loop and the dependence of the associated quark-antiquark potential $V_{q\bar q}$ on the separation $\ell$ between the quarks. Since $V_{q\bar q}\cdot \ell $ is a dimensionless quantity, the potential scales as $V_{q\bar q} \propto \ell^{-1}$ in a CFT. As a consequence, the force between two (infinitely massive) static quarks must scale as
\begin{equation}
\label{eq:scaling_force}
F_{q\bar q} \propto \frac{1}{\ell^2}
\end{equation}
or, equivalently, $F_{q\bar q}\cdot \ell^2$ is dimensionless and independent of $\ell$. Therefore, we can estimate the width of the walking region by examining the range of separations in which this quantity does not vary with the distance between the quarks, up to some designated  tolerance. Defining a threshold $\nu\ll1$, we check the condition 
\begin{equation}
\label{eq:thereshold}
\left|\frac{\dd \log (F_{q\bar q}\ \ell^2 )}{\dd \log \ell}\right|<\nu\,.
\end{equation}
From this we get $\ell_{\text{\tiny UV}}$ and $\ell_{\text{\tiny IR}}$, and their associated energy scales $\mu_{\text{\tiny UV}}$ and $\mu_{\text{\tiny IR}}$, between which the theory is approximately a CFT, as in Fig.~\ref{fig:walkingregime}. The technical details of the computation of the holographic Wilson loop and the quark-antiquark potential can be found in Appendix \ref{sec:ED}.

\begin{figure}[t!]
	\begin{center}
		\includegraphics[width=.80\textwidth]{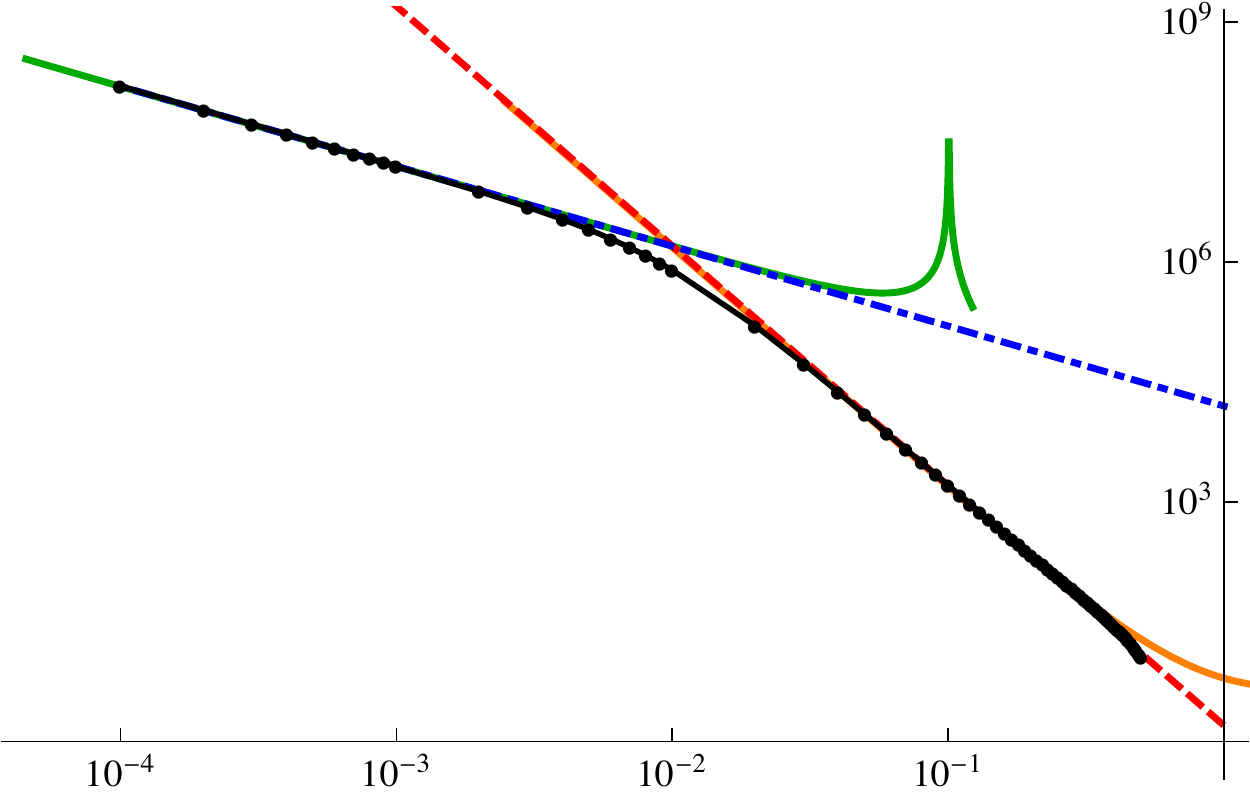}
		\put(-355, 28){$  \epsilon $}
		\put(-65, 220){$\log\frac{\mu_{\text{\tiny UV}}}{\mu_{\text{\tiny IR}}}$} \\[15mm]
		\includegraphics[width=.80\textwidth]{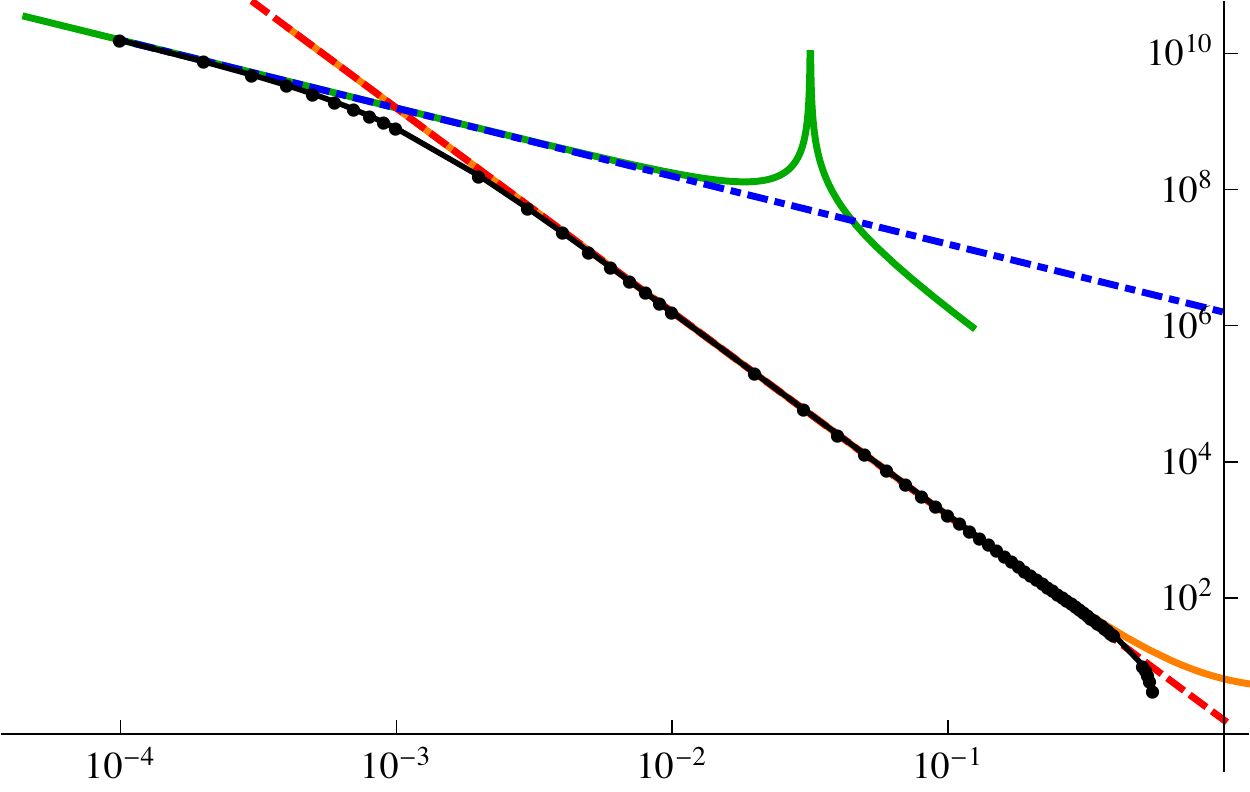}
		\put(-355, 28){$  \epsilon $}
		\put(-65, 220){$ \log\frac{\mu_{\text{\tiny UV}}}{\mu_{\text{\tiny IR}}} $}
		\caption{\small Scalings of the width of the walking region for different values of $\epsilon$ and the different approximations when $\delta=10^{-2}$ (top) and $\delta=10^{-3}$ (bottom). In black we show the data found numerically, the blue dashed-dotted line corresponds to $\pi/(2\epsilon\delta^2)$, the green curve corresponds to $4\pi/|\text{Im}\Delta|$, the red dashed line corresponds to 
			$181\pi/ (360\epsilon^3)$, and the orange curve corresponds to 
			$2\pi \, |\text{Im}\Delta_- + \text{Im}\Delta_+| / (\text{Re }\gamma)^2$, 
			as predicted by \eqref{eq:newscaling}.}
		\label{fig:delta10m2}
	\end{center}
\end{figure}

\begin{figure}[t]
	\begin{center}
		\begin{subfigure}{.48\textwidth}
			\includegraphics[width=1\textwidth]{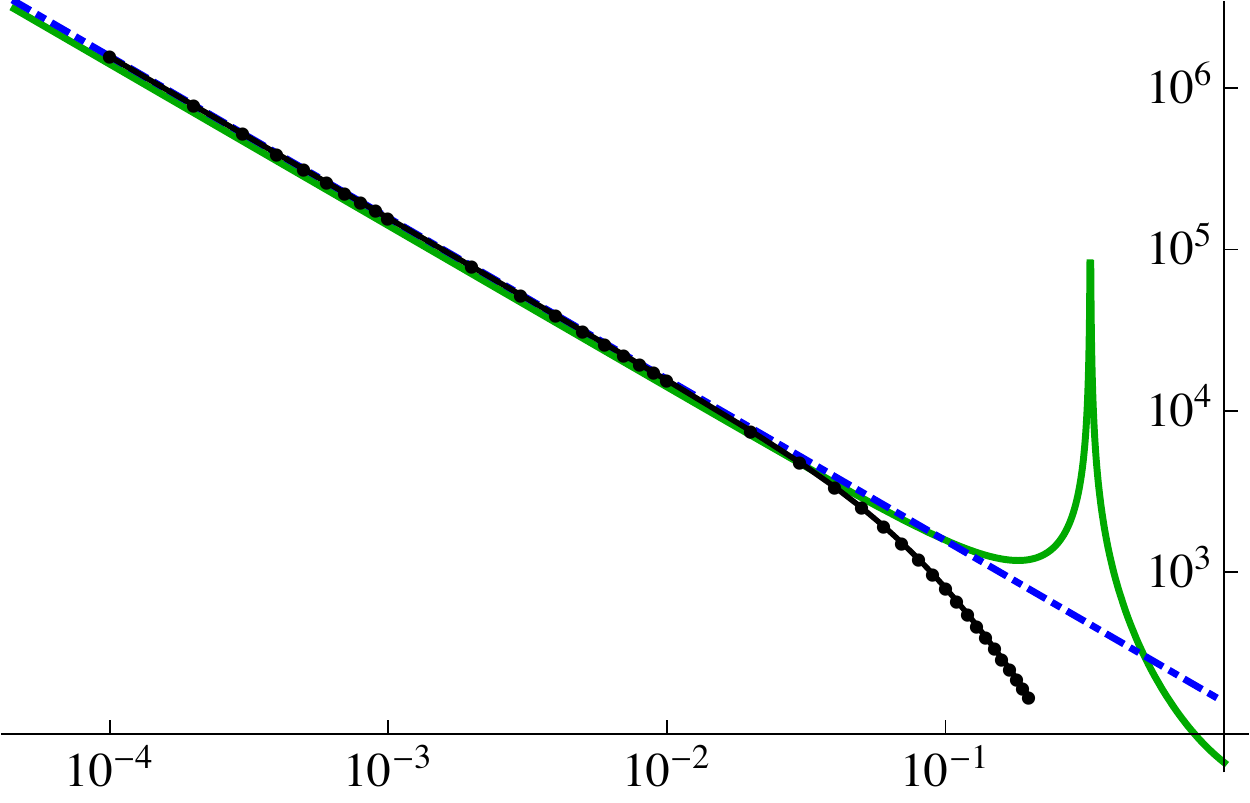}
			\put(-210, 15){$  \epsilon $}
			\put(-50, 140){$\log\frac{\mu_{\text{\tiny UV}}}{\mu_{\text{\tiny IR}}}$}
		\end{subfigure}
		\hfill
		\begin{subfigure}{.48\textwidth}
			\includegraphics[width=1\textwidth]{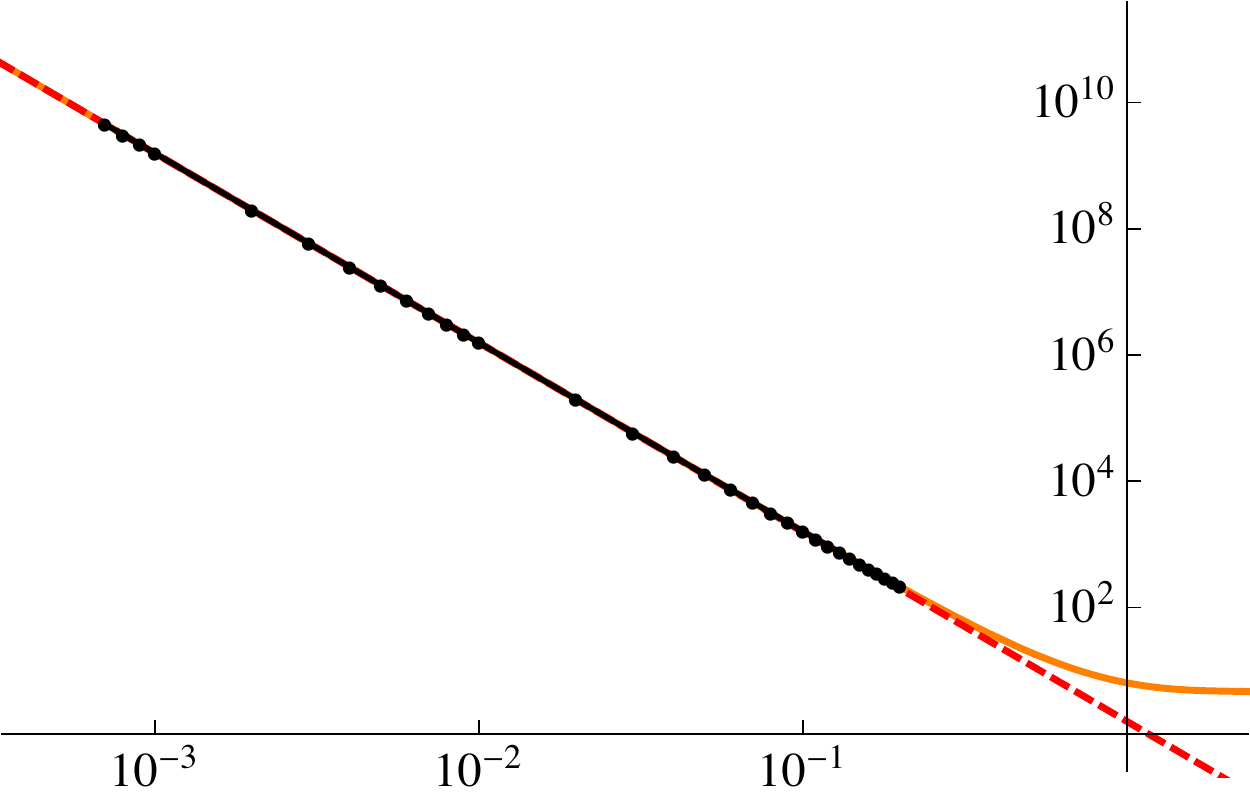}
			\put(-210, 15){$  \epsilon $}
			\put(-60, 140){$\log\frac{\mu_{\text{\tiny UV}}}{\mu_{\text{\tiny IR}}}$}
		\end{subfigure}
		\caption{\small Same color-coding as in Fig.~\ref{fig:delta10m2} for the values $\delta = 10^{-1}$ (left) and $\delta=10^{-7}$ (right). The expected behaviours \eqref{eq:twice} and \eqref{eq:newscaling} for $\epsilon\ll \delta<1 $ and $\delta\ll\epsilon$ are respectively recovered.}
		\label{fig:ScalingsDelta1and7}
	\end{center}
\end{figure}

We fixed $d=4$, the UV dimension to $\Delta_{\text{\tiny UV}}=3$, the distance between the real FP and the complex ones to $\phi_0=1$ and explored the space of parameters $\{\epsilon,\delta\}$ to analyze how they influence the scaling properties of the model. First, tuning the separation between pairs to $\delta=10^{-2}$ we analyzed the dependence of the walking region on the distance to the real axis $\epsilon$. The results are summarized in Fig.~\ref{fig:delta10m2}(top). For values $\epsilon\ll\delta$ our results show that the size of the walking region grows parametrically as $1/(\epsilon\delta^2)$ which, when written in terms of the conformal dimension, matches precisely Eq.~\eqref{eq:twice}. This demonstrates that we are capturing the effects of both pairs of cFPs, with both walking regions overlapping. On the other hand, enlarging $\epsilon$ so that it is still small but $\delta\ll\epsilon$, the system transitions to a regime where the hierarchy grows as $1/\epsilon^3$. This is the enhancement we found in the general analysis when the two pairs of complex fixed points are close. The scaling dictated by Eq.~\eqref{eq:newscaling} is nicely reproduced. 

The computation is repeated for $\delta=10^{-3}$ with the output displayed in Fig.~\ref{fig:delta10m2}(bottom). The results are completely equivalent, but the region in which the regime $\delta\ll \epsilon<1$ is satisfied is larger so the scaling given by Eq.~\eqref{eq:newscaling} is valid within a wider range of $\epsilon$.
Finally, we can magnify the values of $\delta$ so that both regimes can be seen separately in two different cases. This is reflected in Fig.~\ref{fig:ScalingsDelta1and7}, for which the parameters $\delta = 10^{-1}$ and $\delta=10^{-7}$ are chosen in order to explore the regimes $\epsilon\ll\delta$ and  $\delta\ll\epsilon$ respectively.


\section{Complex RG flows}
\label{sec:RGflows}

\subsection{RG flows between complex fixed points}

In this section we explore the structure of RG flows between complex fixed points, which have some peculiar features when compared to real ones. These are mostly due to the fact that the scaling dimension of the operator driving the flow generically has an imaginary part. Indeed, the real part of the dimension determines the nature of the fixed point, i.e., if it is repulsive (the operator is relevant, $\operatorname{Re}\Delta<d$) or attractive (the operator is irrelevant, $\operatorname{Re}\Delta>d$), while the imaginary part is responsible for oscillations around this behavior. This produces a characteristic spiraling structure, as was shown in \cite{Faedo:2019nxw} for a simple holographic example flowing from a real to a complex fixed point. 

Some novel effects appear in the presence of additional complex fixed points. The holographic model of \Sec{sec:strong} is suitable for studying RG flows between different pairs. Since the scalar $\phi$ is dual to the coupling in the field theory, we need to solve for it in terms of the radial coordinate, that is, the energy scale. As we mentioned earlier, for simplicity we will restrict our attention to RG flows that can be found by solving the first-order equations \eqref{eq:BPS} that follow from the superpotential of the theory \eqref{eq:derivativeW}. These are a subclass of the entire set of solutions to the second-order equations of motion and they will suffice to illustrate the physics that we are interested in.

In order to capture RG flows involving cFPs we extend the scalar and the warp factor to complex values and split the equations into real and imaginary parts. It is then simple to solve for the behavior of the real and imaginary parts of the scalar along the flow. For additional details see \cite{Faedo:2019nxw}. 

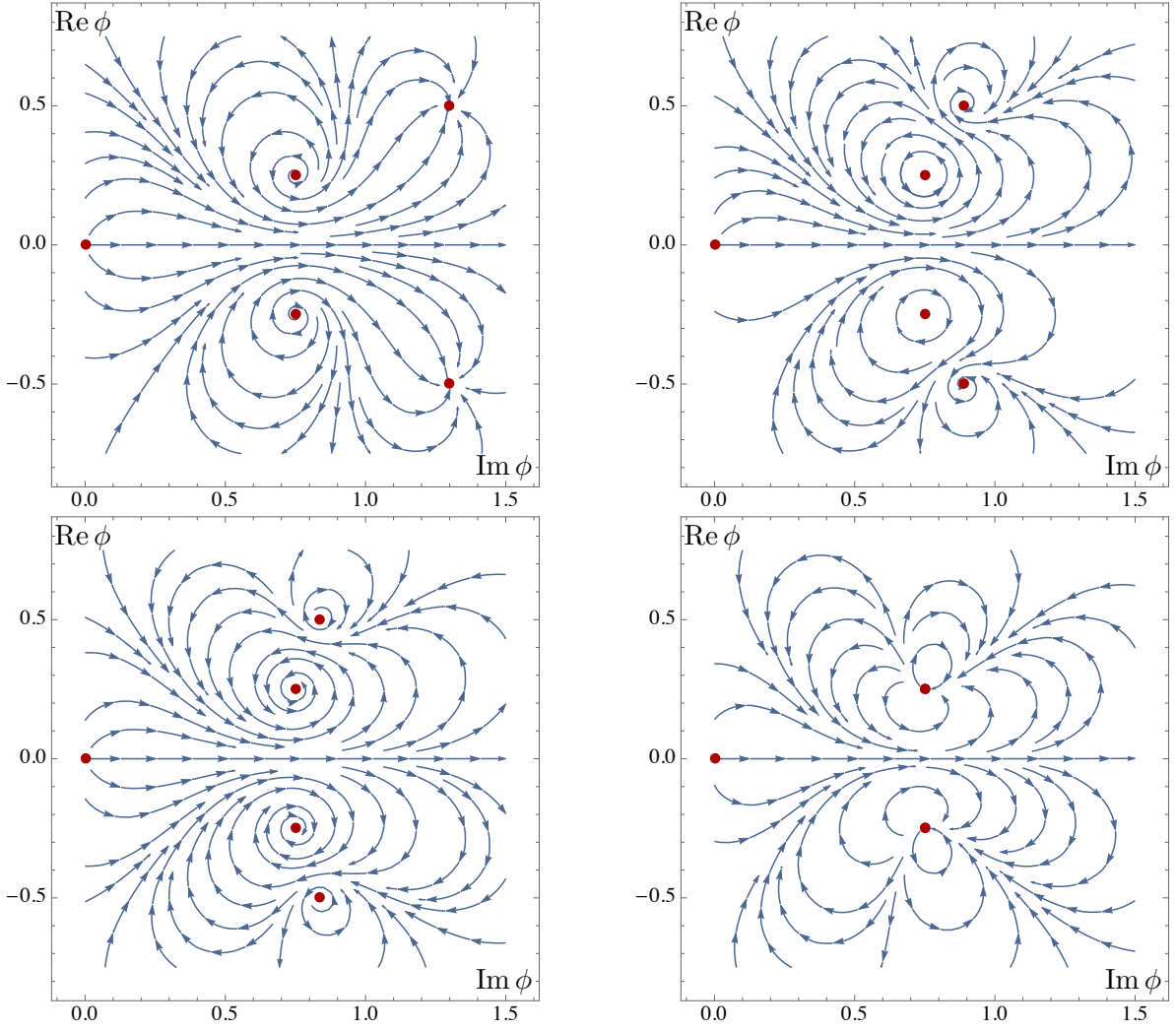
\begin{figure}[h!!!]
	\begin{center}
		\begin{subfigure}{.46\textwidth}
			\includegraphics[width=1\textwidth]{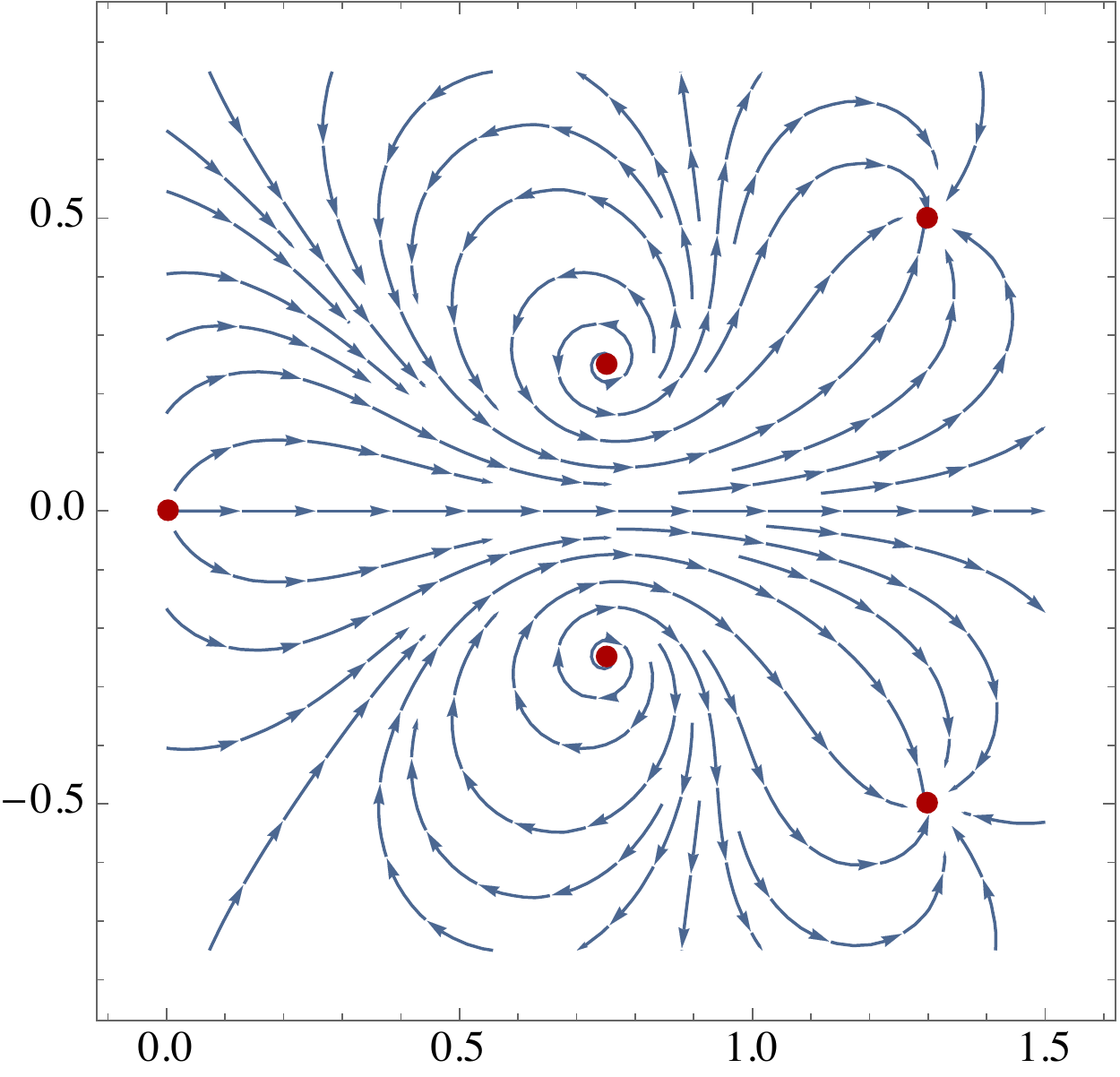}
			\put(-25, 15){$ \operatorname{Im}\phi $}
			\put(-185, 185){$ \operatorname{Re}\phi $}
		\end{subfigure}
		\hfill
		\begin{subfigure}{.46\textwidth}
			\includegraphics[width=1\textwidth]{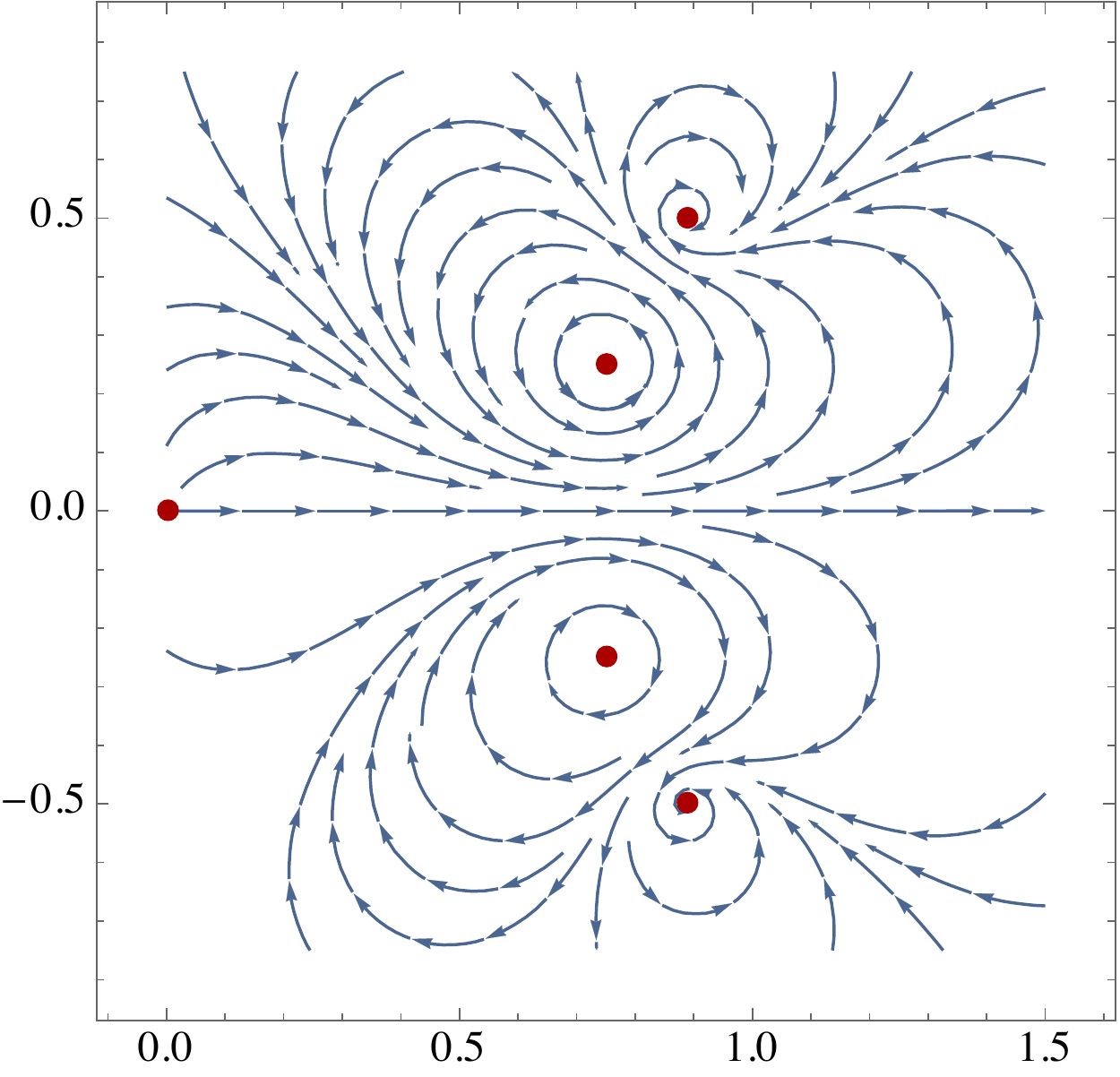}
			\put(-25, 15){$ \operatorname{Im}\phi $}
			\put(-185, 185){$ \operatorname{Re}\phi $}
		\end{subfigure}
		\begin{subfigure}{.46\textwidth}
			\includegraphics[width=1\textwidth]{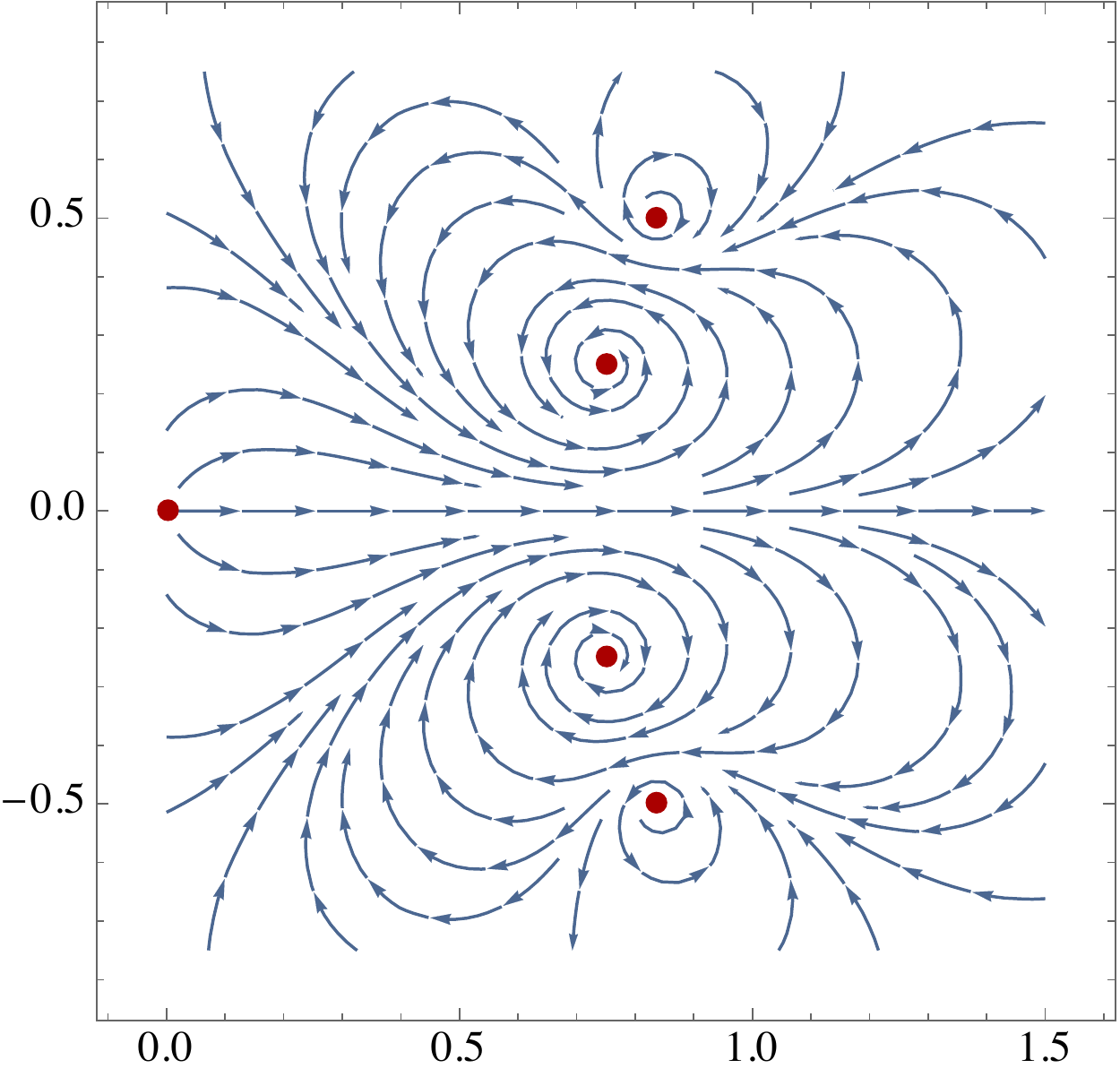}
			\put(-25, 15){$ \operatorname{Im}\phi $}
			\put(-185, 185){$ \operatorname{Re}\phi $}
		\end{subfigure}
		\hfill
		\begin{subfigure}{.46\textwidth}
			\includegraphics[width=1\textwidth]{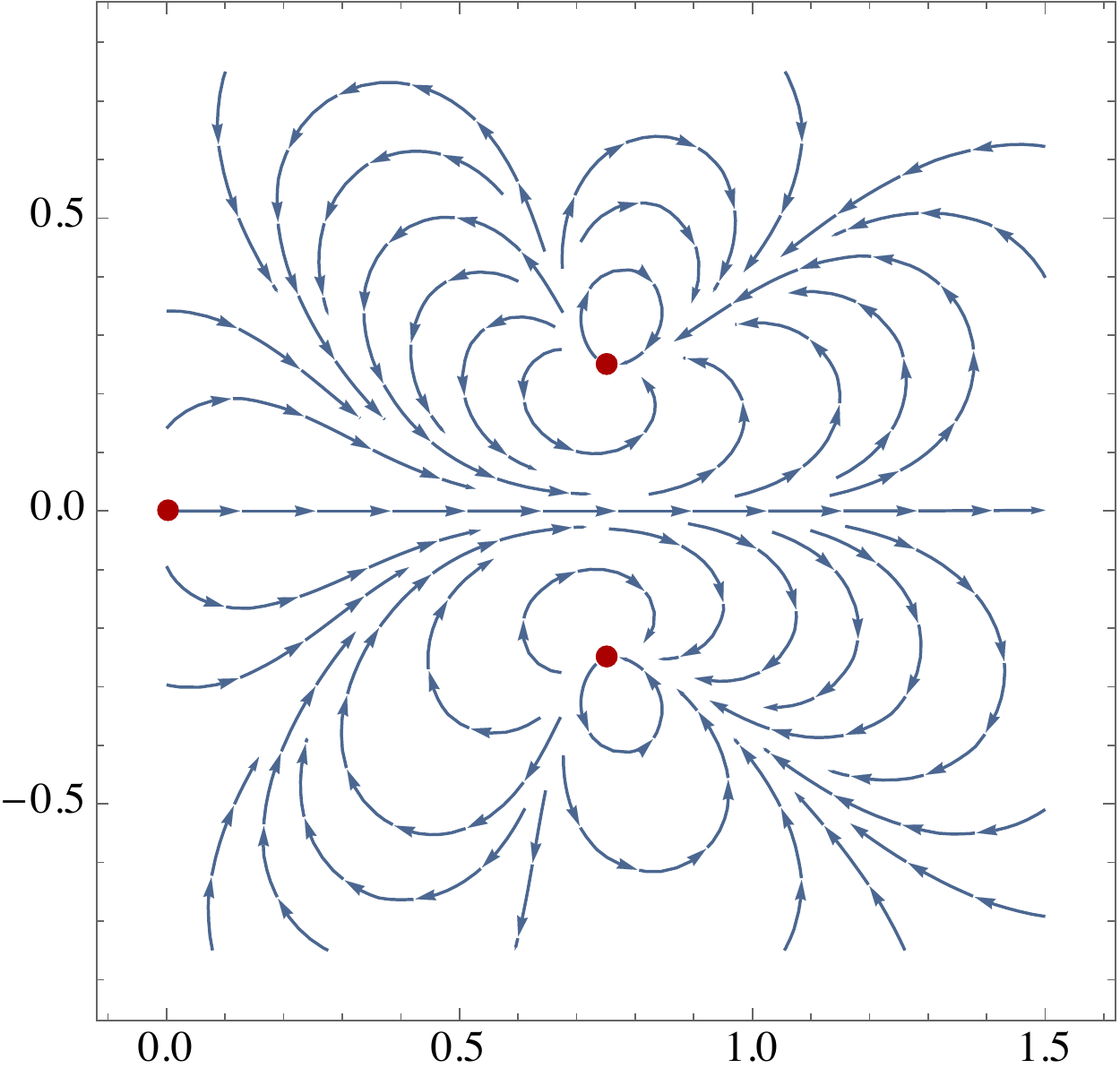}
			\put(-25, 15){$ \operatorname{Im}\phi $}
			\put(-185, 185){$ \operatorname{Re}\phi $}
		\end{subfigure}
		\caption{\small Sequence of RG flows between the different fixed points in the model \eqref{eq:derivativeW}. The first pair is located at $\phi_1=\bar{\phi}_1^*=0.75+0.25i$ while the location of the second one varies. The top-left panel shows the typical flow when the pairs of cFPs are widely separated, the first pair being repulsive and the second one attractive. As they come closer, there is a particular separation at which the scaling dimension at the first pair verifies $\operatorname{Re}\Delta=d$ but $\operatorname{Im}\Delta\ne0$, resulting in closed flows encircling it. Passing that separation, the nature of the fixed points is interchanged and the second pair becomes repulsive. Finally, when the two pairs of cFPs merge the operator becomes marginal, so the flows are closed, starting and ending at the same fixed point.}
		\label{fig:cRGflows}
	\end{center}
\end{figure}

A typical example when the pairs are widely separated is provided in the top-left panel of Fig.~\ref{fig:cRGflows} where, together with the flows between the real UV fixed point and the second pair of cFPs, we observe spiraling flows between the first and the second pair of cFPs. Note that the unique purely real flow, the horizontal one passing exactly in between both pairs of complex fixed points, displays some sort of scaling if these are close enough to the real axis. 

The conformal dimension of the operator depends on the distance between the complex fixed points. As they come closer, there is a special value at which $\operatorname{Re}\Delta=d$ but $\operatorname{Im}\Delta\ne0$ at the first pair. The operator is thus neither repulsive nor attractive, so the RG flows do not start or end at the fixed points. Nevertheless, the imaginary part produces non-trivial solutions that encircle the fixed point, seen in the top-right panel of the figure. This critical value is also the boundary at which the nature of the pairs is interchanged, the first one becoming attractive and the second one repulsive, as in the bottom-left panel. Finally, when the fixed points coalesce, the operator becomes marginal. Still, the higher order corrections produce an RG flow that both starts and ends at the same fixed point. Although this is reminiscent of  the so-called ``boomerang flows'' \cite{Donos:2017ljs,Donos:2017sba},  an important difference is that the flows considered here are Lorentz-invariant. Note that at this type of fixed point the operator associated to the direction of the boomerang flows can be both relevant or irrelevant depending on the side of the fixed point we are at. As the relevant or irrelevant nature of the operator at the IR fixed point (which in this case is the same as the UV fixed point) depends on the value of the coupling, this is reminiscent of the  ``dangerously irrelevant'' operators \cite{Amit:1982az,Gukov:2015qea}. In principle a similar behavior could be observed for fixed points annihilating on the real axis.

If the coupling is separated into its real and imaginary parts, the complex RG flows can be interpreted as flows in a two-dimensional dynamical system. It may be possible to use this viewpoint to classify complex fixed points and flows generalizing the proposal of \cite{Gukov:2016tnp} for real couplings.

\subsection{On the $c$-theorem}

At complex fixed points the central charge (usually proportional to the radius of the dual AdS solution) is in general a complex number. Therefore it does not make sense to ask whether this quantity increases or decreases along a complex RG flow between fixed points. However, one may wonder whether some sort of $c$-theorem may still apply, meaning that perhaps the real part or the modulus of the central charge could exhibit some monotonic behaviour. We will see in this section that the answer to this question is negative. The reason behind it is presumably that complex CFTs are not unitary, which is one of the assumptions of the theorem.

 A simple holographic way to see that no version of the $c$-theorem holds  is the following. In holography the $c$-function is defined in terms of the warp function in domain wall coordinates \eqref{eq:metric} as
\begin{equation}
c\propto\frac{1}{\left(A'\right)^3}\,,\qquad\Rightarrow\qquad c'\propto-\frac{A''}{\left(A'\right)^4}\,,
\end{equation}
with both proportionality constants being positive. The derivative is taken with respect to the radial coordinate, which as usual plays the role of the energy scale in the field theory dual. This gives, at the fixed points, $c\propto L^3$, with $L$ the AdS radius. Considering a single scalar (as in Eq. \eqref{eq:action}) with radial dependence, one of Einstein's equations states that 
\begin{equation}\label{eq:App}
A''\,=\,-\frac{1}{2\left(d-1\right)}\left(\phi' \right)^2\le0\,,
\end{equation}
which shows the monotonicity property
\be
c'\ge0\,.
\ee
Therefore the holographic $c$-function decreases towards the IR. In terms of the gravitational dual this means that the radii of AdS solutions must satisfy $L_{\text{\tiny UV}}>L_{\text{\tiny IR}}$.

As in the previous section, to accommodate cFPs we promote the scalar, and for consistency the warp factor $A$, to complex functions with both real and imaginary parts. Thus, Eq. \eqref{eq:App} splits into real and imaginary parts as 
\be
\begin{aligned}\label{eq:appcomplex}
A''_{\rm R}&=-\frac{1}{2\left(d-1\right)}\left[\left(\phi'_{\rm R} \right)^2-\left(\phi'_{\rm I} \right)^2\right]\,,\\[2mm]
A''_{\rm I}&=-\frac{1}{\left(d-1\right)}\,\phi'_{\rm R}\, \,\phi'_{\rm I}\,, 
\end{aligned}
\ee
so there are no obvious positivity constraints. As a consequence, if one identifies the central charge at the cFP  with the complex AdS radius, it is entirely possible to flow from a cCFT to another one with ``larger $c$'', meaning that both its real part and/or  its modulus can increase along the flow.

We will illustrate this feature with a simple holographic model given by a superpotential with derivative 
\be\label{eq:Wcviolation}
W'\left(\phi\right)=\frac{W_0}{L}\,\phi\left(\phi-\phi_1\right)\left(\phi-\overline{\phi}_1\right)\left(\phi-\phi_2\right)
\ee
where, as in the previous section, the constant $W_0$ can be related to the dimension of the dual operator at the UV fixed point $\phi=0$. In the following we fix it so that $d=4$ and $\Delta_{\text{\tiny UV}}=3$ at the UV with radius $L$. Additionally, it has a pair of complex fixed points located at $\phi=\phi_1$ and its conjugate and another real fixed point at $\phi=\phi_2$. This second real fixed point is always attractive, so it corresponds to an IR CFT. The nature of the complex fixed points depends on the parameters of the model and can be both repulsive or attractive, with flows either starting or ending on them. In this way, this superpotential captures the physics of the $\beta$-function \eqref{eq:spoil}.

\begin{figure}[t]
	\begin{center}
			\includegraphics[width=0.50\textwidth]{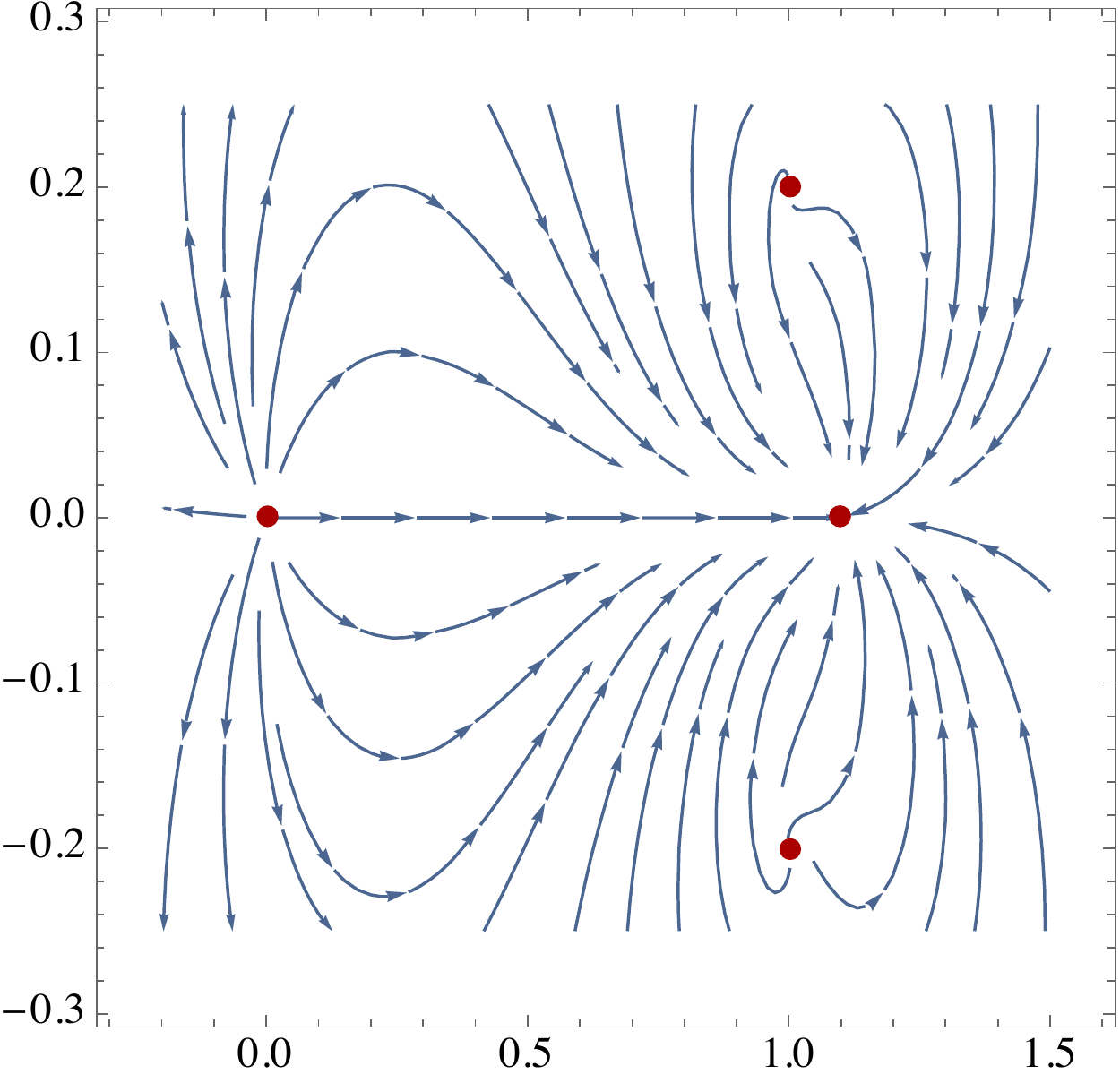} 
			\put(-210,220){$\operatorname{Im}\phi $}
			\put(5,10){$ \operatorname{Re}\phi$}
		\caption{\small  Example of complex RG flows for $d=4$ violating the $c$-theorem. The flow from $\phi=0$ is triggered by a source for a $\Delta_{\text{\tiny UV}}=3$ operator and the cFPs are located at $\phi_1=\overline{\phi}_1^*=1+0.2i$ while the real one is at $\phi_2=1.1$. The ratio of their respective AdS radii is $L_1/L_2=0.999979-0.000083i$, so the flow between them that can be seen in the figure violates the standard monotonicity theorem.}\label{fig:c-violation}
	\end{center}
\end{figure}

This model contains flows that violate the $c$-theorem. To see this, notice that from the first equation in \eqref{eq:appcomplex} the condition $A''_{\rm R}\le0$ is more likely to be violated when the gradient of the imaginary part of the scalar is larger than its real counterpart. This means that flows between $\phi_1$ and $\phi_2$ in a model in which $\phi_2$ is close to the real part of $\phi_1$, and therefore $\phi_R$ barely changes along the flow, are good candidates for the possible violation of the theorem. Indeed, it can be seen that when ${\rm Re}\,\phi_1=\phi_2$ the AdS radius at $\phi_1$ is smaller than that at $\phi_2$ (both its real part and its modulus) and nevertheless the operator at $\phi_1$ is relevant, so there is a flow that starts from it and ends at $\phi_2$, violating the $c$-theorem (it cannot end at $\phi=0$ because this fixed point is always an UV one). This is also true on an interval of order ${\rm Im}\,\phi_1$ around that value. An example is given in Fig.~\ref{fig:c-violation}. Interestingly, this type of RG flows occur in a region of parameter space in which there is no hierarchy of scales due to closeness of the real fixed point to the complex ones, as explained at the end of \Sec{sec:toy}. 

\section{Discussion}

Fixed point annihilation is the mechanism behind phenomena like walking in gauge theories \cite{Kaplan:2009kr} or weak first-order phase transitions in condensed matter physics \cite{Gorbenko:2018ncu, Gorbenko:2018dtm}. For it to take place there must be a pair of real fixed points of the $\beta$-function that, by tuning some external parameter, can be made to merge and become a pair of complex conjugate fixed points. When these are close to the real axis, the RG flow slows down while passing in between them, resulting in a large hierarchy of scales dictated by Miransky scaling. A useful invariant characterization of this scaling can be given in terms of the conformal dimension of the operator driving the flow referred to the putative cCFT that can be defined at a cFP \cite{Gorbenko:2018ncu, Gorbenko:2018dtm}, as seen in Eq.~\eqref{eq:Miransky}.

In this work we have generalized this picture to the case of several complex fixed points and observed an interesting non-trivial interplay between the different pairs. We have shown that the way in which the fixed points are distributed in the complex plane has an impact on the properties of the RG flow. If two pairs of complex fixed points are close to the real axis but widely separated there is no influence, as expected. There is a walking region in the vicinity of the first pair followed by a transition period and another quasi-conformal behavior when the second pair is reached. 

Both walking regions start to overlap as the pairs of cFPs become closer and as a result the hierarchy of scales widens, while still being parametrically of 
$O\left(\exp \epsilon^{-1}\right)$, with $\epsilon$ the distance of the pair to the real axis. However, when the distance $\delta$ between pairs is $O\left(\epsilon\right)$ or smaller there is an enhancement of the effect and the scaling becomes 
$O\left(\exp \epsilon^{-3}\right)$. Moreover, Miransky scaling takes a different form when written in terms of conformal data, as can be seen for instance in 
Eqs.~\eqref{eq:scalingC} and \eqref{eq:newscaling}.  

The precise form of the scaling can be model dependent, but the enhancement seems to be a universal feature of models with several pairs of complex fixed points if these are close enough. We have checked this both at weak coupling using toy-model $\beta$-functions and explicit field theory examples as well as at strong coupling by means of the holographic duals proposed in \cite{Faedo:2019nxw}. In all these cases the diagnosis for this effect is the presence of a marginal operator in the complex conformal field theory. 

This enhancement may have interesting phenomenological consequences. For instance, in the explicit field theory models of \Sec{sec:FT} it generates a hierarchy of hierarchies between the different couplings in the sense that, even if both walk in some energy interval, the range in which the marginal coupling (in the cCFT sense) behaves in a quasi-conformal way is parametrically larger. It would be interesting to find other explicit examples of this behavior in the large collection of $\beta$-functions that are available in the literature. 

While the case of a single coupling with several pairs of complex fixed points has essentially been covered in detail in this work, models with more than one coupling have a large number of alternatives that could include new interesting effects. For instance, they may include not only points but entire lines of complex fixed points, possibly coming from the annihilation of real ones, an effect that as far as we know has not been explored. It may be worth analyzing at least the general case of two couplings, possibly along the lines of \cite{Gukov:2016tnp}. The simplifications enjoyed by the $\beta$-function of double-trace operators at large-$N$ \cite{Pomoni:2008de} could be useful in this regard. 

The bottom-up holographic implementation suggested in \cite{Faedo:2019nxw} captures correctly the expected behavior. It also allowed us to study renormalization group flows between complex fixed points, which have peculiar properties. There are closed trajectories encircling the cFPs as well as flows starting and ending in the same cCFT. It is therefore not surprising that the $c$-theorem can be violated.

It would be interesting to find examples of cFPs  in fully-fledged string theory models.


\section*{Acknowledgements}
 
A.F.~and C.H.~have been partially supported by the Spanish {\em Ministerio de Ciencia, Innovaci\'on y Universidades} through the grant PGC2018-096894-B-100. A.F.~is also supported by the ``Beatriz Galindo'' program, reference BEAGAL 18/00222. D.M.~and J.G.S.~are supported by grants SGR-2017-754 and PID2019-105614GB-C22, and they acknowledge financial support from the State Agency for Research of the Spanish Ministry of Science and Innovation through the ``Unit of Excellence Mar\'\i a de Maeztu 2020-2023'' award to the Institute of Cosmos Sciences (CEX2019-000918-M). J.G.S.~is also supported by the FPU program, fellowship FPU15/02551.

\appendix

\section{Arbitrary number of complex fixed points}
\label{sec:arbitrary}

Consider the general case in which the $\beta$-function is a polynomial\footnote{We comment on even more general cases at the end of this Appendix.} and thus can be written as
\begin{equation}\label{eq:betaarb}
\begin{aligned}
\beta &= \prod_{k=1}^{N} (\lambda-\lambda_k) = 
\Big[ (\lambda-\delta)^2+\epsilon^2\Big] \Big[ (\lambda+\delta)^2+\epsilon^2\Big] \prod_{k=5}^{N} (\lambda-\lambda_k)\\
&= \Big[ (\lambda-\delta)^2+\epsilon^2\Big] \Big[ (\lambda+\delta)^2+\epsilon^2\Big]  P(\lambda)\,.
\end{aligned}
\end{equation}

We are thus assuming that $\beta$ is a polynomial of degree $N$ and $\lambda_k$ are its zeros. These can be complex as long as they appear in complex-conjugate pairs in such a way that $\beta$ is real. We want to prove in this Appendix that in this general case it is still satisfied 
\begin{equation}
\begin{aligned}
\log\frac{\mu_{\text{\tiny UV}}}{\mu_{\text{\tiny IR}}}&\propto\frac{1}{\epsilon^3}\quad\qquad\,\, \,\, \text{if }\delta\ll \epsilon \\[3mm]
\log\frac{\mu_{\text{\tiny UV}}}{\mu_{\text{\tiny IR}}}&\propto\frac{1}{\delta^2\, \epsilon}\quad\qquad  \text{if }\delta\gg \epsilon \\
\end{aligned}
\end{equation}
even though the precise coefficients, whose expression we will determine, are model dependent. The argument goes as follows. First, recall that the ratio between the scales is given by
\begin{equation}\label{eq:generalBetaScaling}
\log\frac{\mu_{\text{\tiny UV}}}{\mu_{\text{\tiny IR}}} = \left|\int_{-a}^a \frac{\dd\lambda}{\beta(\lambda)}\right| \,.
\end{equation}
\begin{figure}[t]
	\begin{center}
		\includegraphics[width=.85\textwidth]{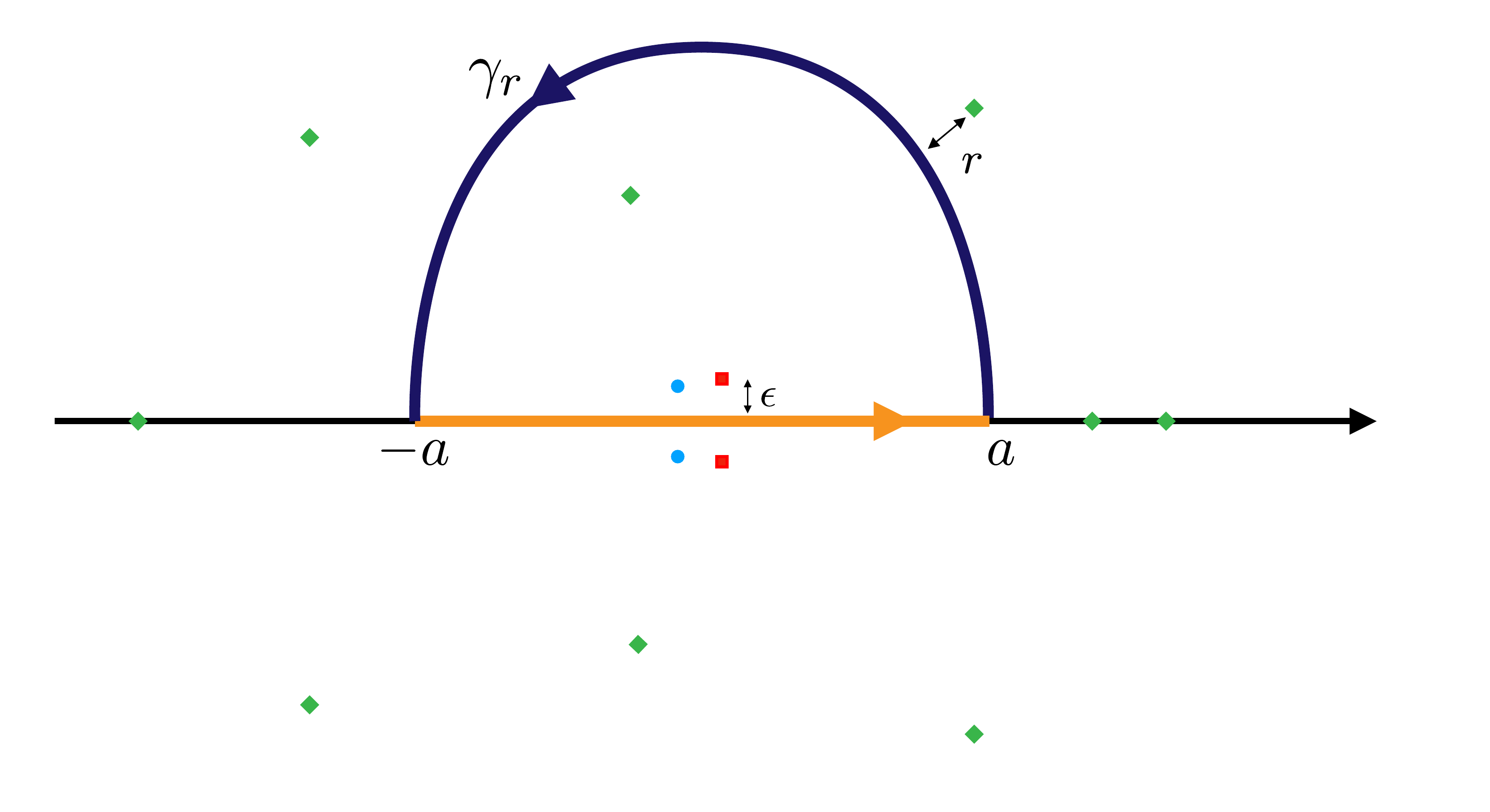}
		\caption{\small Countour integral we are performing. Red squares and cyan dots represent two pairs of cFPs which approach and collide as in Fig. \ref{fig:collision}. Green diamonds represent the zeros of the polynomial $P(\lambda)$. The path $\gamma_r$ is plotted in blue, whereas the interval $[-a,a]$ is plotted in orange. The integral is done anticlockwise. The bound $r$ is given by the zero of $P(\lambda)$ which is closer to the path.}
		\label{fig:complex}
	\end{center}
\end{figure}

Consider a closed path $\Gamma = \gamma_r\circ \gamma_a$, where  $\gamma_r\subset\mathset{C}$ is a path in the complex plane and $\gamma_a$ is the path that extends along the interval $[-a,a]\subset \mathset{R}$ (see Fig.~\ref{fig:complex}). The paths we are choosing are such that:
\begin{itemize}
	\item[(i)] The length of the interval is larger than $\epsilon$ and $\delta$, that is to say, $a \gg \epsilon$ and $a\gg\delta$.
	\item[(ii)] None of the zeros of the beta function belongs to the closed path $\Gamma$, $\lambda_k\notin \Gamma$.
	\item[(iii)] The path $\gamma_r$ does not cross itself nor the real axis.
	\item[(iv)] The path $\gamma_r$ is kept fixed as we vary $\epsilon$ and $\delta$.
	\item[(v)] The modulus of all the points $z\in \gamma_r$ is bounded from below by $|z|>a/2$.
\end{itemize}
Then, we define $r$ as the distance between $\gamma_r$ and the zero of the beta function that is closer to it. Mathematically this means
\begin{equation}
0 < r = \min_{z\in \gamma_r} |z-\lambda_k|,\ {k=1,\cdots ,N}.
\end{equation}
Since $\gamma_r$ is compact and there is a finite number of zeros in $\beta$, the minimum exists and is greater than zero. Additionally, assumptions (iv) and (v) ensure that $r$ does not scale with $\epsilon$ and $\delta$. Therefore 
\begin{equation}\label{eq:boundIr}
|I_r| \equiv \left|\int_{\gamma_r}\frac{\dd z}{\beta(z)}\right| \ =\  \left|\  \int_{\gamma_r}\frac{\dd z}{\prod_{k=1}^N(z-\lambda_k)}\ \right|\  \leq\  \text{length}(\gamma_r)\cdot  r^{-N} 
\end{equation}
Importantly, the modulus of $I_r$ does not scale with $\epsilon$. Then, using the residue theorem
\begin{equation}\label{eq:residueTheorem}
\int_{\Gamma} \frac{\dd z}{\beta(z)} = I_r + \int_{-a}^{a} \frac{\dd \lambda}{\beta(\lambda)} = 2\pi i \sum_{\lambda_k\, \in\, \text{Int}\Gamma}\text{Res}\left(\frac{1}{\beta(z)},\lambda_k\right)
\end{equation}
Taking into account \eqref{eq:boundIr} and that
\begin{equation}
\text{Res}\left(\frac{1}{\beta(z)},\lambda_k\right) = O (\epsilon^0) \qquad k =5,\cdots ,\ N
\end{equation}
we conclude from \eqref{eq:residueTheorem} that the leading-order-in-$\epsilon$ contribution to \eqref{eq:generalBetaScaling} comes from the two poles which are merging with positive imaginary part
\begin{equation}
\label{eq:residueAtEpsilon}
\text{Res}\left({\beta(z)}, - \delta + i\epsilon\right) +\text{Res}\left({\beta(z)},\delta + i\epsilon\right) \,,
\end{equation}
leading to
\begin{equation}
\begin{aligned}
\label{eq:scalinggeneral}
\int_{-a}^a \frac{\dd \lambda}{\beta(\lambda)} &\approx \frac{\pi}{2\epsilon^3}\cdot \frac{1}{P(0)}\quad \qquad\,\,\, \delta\ll \epsilon\\[3mm]
\int_{-a}^a \frac{\dd \lambda}{\beta(\lambda)} &\approx \frac{\pi}{2\delta^2\epsilon }\cdot \frac{1}{P(0)} \quad\qquad \epsilon\ll \delta
\end{aligned}
\end{equation}

Let us now try to find the scaling in terms of the real and imaginary parts of $\Delta$. In contrast to the simple model where the $\beta$-function admitted only of the four colliding fixed points (\textit{i.e} when $P(\lambda)=1$), in the general case when $\delta\ll \epsilon$ we will need to distinguish between the dimensions of the operator in the two different cCFTs located at $\lambda \, =\,  \pm\, \delta +i\, \epsilon $, which we denote by $\Delta_\pm$. These are given by
\begin{eqnarray}
\text{Re}\, (\,d-\Delta_\pm) & =& \Big( \mp \, 8 P(0) \, \epsilon^2\, +\, O(\epsilon^3) \Big)\, \delta\ +\ O (\delta^2)\\[2mm]
\text{Im}\ \Delta_\pm &= &  \Big( \mp \, 8 P'(0)\, \epsilon^3\, +\, O(\epsilon^4) \Big)\, \delta  \, + \Big( 8 P(0) \epsilon\ + \ O (\epsilon^2)\Big)\, \delta^2\ + \ O(\delta^3)\nonumber
\end{eqnarray}
The leading term in $ \text{Im}\ \Delta_\pm $ depends on whether $P'(0)$ vanishes (as in Sec. \ref{sec:toy}) or not. If it is nonzero, expressing the scaling in terms of a single dimension $\Delta_+$ or $\Delta_-$ gives rise to complicated formulas with explicit dependence on $\beta$-function parameters. However, notice that the combination
\begin{equation}
\text{Im}\, \Delta_-   +\  \text{Im}\, \Delta_+ = \Big( 16P(0)\, \epsilon \, + \, O(\epsilon^2) \Big) \, \delta^2 + O(\delta^3)
\end{equation}
allows us to write the walking region as
\begin{eqnarray}
\int_{-a}^a \frac{\dd \lambda}{\beta(\lambda)} &\approx&\frac{2\pi}{\text{Re}(d-\Delta_\pm)^2}\, \Big(\text{Im}\,\Delta_-+ \text{Im}\,\Delta_+\Big)
\end{eqnarray}
Indeed, when $\Delta_+ = \Delta_-$ this result agrees with \eqref{eq:scalingC}.

On the other hand, when $\epsilon\ll\delta$ the scaling is simply
\begin{equation}
\int_{-a}^a \frac{\dd \lambda}{\beta(\lambda)} \ \approx\ \frac{4\pi}{| \text{Im}\ \Delta_\pm|}\qquad \epsilon\ll\delta\ \,,
\end{equation}
which is again twice the value of the scaling when there is a single pair of fixed points colliding.

Let us finish this Appendix with a final remark. Even though we have restricted ourselves to $\beta$-functions of the form \eqref{eq:betaarb}, similar arguments with additional considerations stand in more general cases. For example, $P(\lambda)$ could be a meromorphic function (in particular, a quotient of polynomials), since its poles would result in zeros of $1/\beta$, which do not affect the reasoning. In the case of an infinite number of zeros in $P(\lambda)$, it is in principle still possible to find an appropriate path $\gamma_r$ with the desired properties, given that zeros of holomorphic functions are always isolated. Finally, in a more physical case the location of all the cFPs may vary as the parameters $\epsilon$ and $\delta$ change. In such scenario one might still apply the same arguments if allowing \textit{small} deformations of $\gamma_r$ with $\epsilon$ and $\delta$ permits to find some bound like \eqref{eq:boundIr}.

\section{Holographic Wilson loops for cCFTs}
\label{sec:ED}

In this Appendix we provide the details of the computation of the holographic Wilson loop and the associated quark-antiquark potential. Consider a gravitational action of the form \eqref{eq:action} and suppose that it is characterized by a superpotential that can be written as
\begin{equation}\label{eq:superpotential}
\begin{aligned}
	W(\phi) = -\frac{1}{L}\left(1+\frac{1}{12}\phi^2 + W_{\text{\tiny H}}(\phi)\,\,\phi^3\right)\,.
\end{aligned}
\end{equation}
We take this as a definition of $W_{\text{\tiny H}}(\phi)$. From this one can obtain the following potential 
\begin{equation}\label{eq:potential}
V(\phi) = (d-1)\left[ 2(d-1)\left(\frac{\dd W}{\dd \phi}\right)^2 - d \,\, W^2\right] =\frac{1}{L^2}\left[-12-\frac{3}{2}\phi^2 +O(\phi^4)\right]\,,
\end{equation}
where we have fixed the dimension to $d=4$. In this way, the scalar has mass $m^2L^2=-3$ around the AdS geometry at $\phi=0$ and thus the operator triggering the flow from the UV has conformal dimension $\Delta_{\text{\tiny UV}}=3$. The superpotentials used in Secs.~\ref{sec:strong} and \ref{sec:RGflows} for the holographic computations can be written as \eqref{eq:superpotential}.

We write the metric in Domain Wall coordinates as
\begin{equation}
\label{eq:metric}
\dd s^2 = e^{2A}\parent{-\dd t^2 + \dd  \vec{x}^2} + \dd\rho^2=  e^{2A}\parent{-\dd t^2 + \dd  \vec{x}^2} + \rho'(\phi)^2\ \dd\phi^2\,,
\end{equation}
where we are using the scalar itself as radial coordinate in the second equality. The background solutions can be found by solving the following first order equations
\begin{equation}\label{eq:BPS}
\frac{\dd A}{\dd\r} = W\ ,\qquad \frac{\dd \phi}{\dd \r} = -2\ (d-1)\ \frac{\dd W}{\dd \phi}\,,
\end{equation}
which imply the second order ones. In terms of the scalar as radial coordinate the equation for the warp factor reduces to
\begin{equation}\label{eq:Aofphi}
\frac{\dd A}{\dd \phi} = \frac{\dd A}{\dd \rho} \, \parent{\frac{\dd \phi}{\dd\r}}^{-1}= -\frac{1}{2\ (d-1)}\,\, \frac{W}{\dd W/\dd\phi}\,.
\end{equation}
Note that the right hand side of \eqref{eq:Aofphi} is known analytically in our models, so $A$ can be found by a simple (possibly numerical) integration. On the other hand, its UV behaviour can be straightforwardly obtained, giving
\alignedeq{\label{eq:AintheUV}
A &= -\log\phi +A_0 \ +\ 18 \ W_H(0)\  \phi\ + \  O(\phi^2)\,,\\[2mm]
e^A &= \frac{1}{\phi} + 18\ W_H(0)\ + \ O(\phi)\,,
}
where in the second line we have already set the innocuous integration constant to $A_0 = 0$.

To compute the Wilson loop, we consider as usual a string hanging from the boundary and whose endpoints are located $x_1=\pm \frac{\ell}{2}$. Because of symmetry, it has a turning point at $x_1 = 0$ and some value $\rho_*$ of the radial coordinate, corresponding to a scalar and warp factor $\phi=\phi_*$ and $A=A_*$ respectively. The embedding of the string in the background \eqref{eq:metric}, specified by $x_1\left(\rho\right)$, is such that the induced metric reads
\begin{equation}
\label{eq:embeddingstring}
\dd s^2_s = e^{2A}\left(-\dd t^2 \, +\,  (\dot x_1(\rho))^2 \dd\rho^2\right) + \dd\rho^2= -e^{2A}\dd t^2 + \left(1+e^{2A}(\dot x_1)^2\right)\dd \rho^2 \,.
\end{equation}
Using that the integrand is symmetric with respect to $x_1=0$ and integrating the time coordinate the action for the string reads
\alignedeq{\label{eq:WL}
S_{q\bar q} &= -\frac{1}{2\pi\alpha'} \int \dd t\dd x_1 \sqrt{-\det g|_s} =
 - \frac{T}{2\pi\alpha'} \int_{-\frac{\ell}{2}}^{\frac{\ell}{2}} \, \dd \rho \ e^{A}\sqrt{1+e^{2A}(\dot x_1)^2}\,.
}

Note that the action depends on the derivative of the field $x_1$, but not on the field itself. Consequently, the conjugate momentum is constant:
\begin{equation}
\pi_x = \frac{\delta S_{q\bar q}}{\delta \dot x_1} = -\frac{T}{2\pi\alpha'} \, \cdot\, \frac{e^{3A}\, \, \dot x_1}{\left(1+e^{2A}(\dot x_1)^2\right)^{\frac{1}{2}}} =   \frac{T}{2\pi\alpha'} \, p 
\end{equation}
The constant $p$ can be determined by realizing that at the turning point $\rho = \rho_*$, the derivative of the field $x_1$ diverges, leading to
\begin{equation}
p=e^{2A_*} \,.
\end{equation}
With this information we can also write
\begin{equation}
\dot x_1 (\rho) = - \frac{e^{2A_*-A}}{\sqrt{e^{4A}-e^{4A_*}}},
\end{equation}
where we are considering the branch where $x_1(\rho)\in [-\ell/2,0]$. This allows us to express the separation between the two quarks for each choice of the position of the turning point $\rho=\rho_*$ in terms of the simple integral
\begin{equation}\label{eq:Lfromphi}
\ell= 2 \int_{0}^{\frac{\ell}{2}} \dd x_1 = 2\int_{\rho_*}^{\infty} \dd\rho \frac{e^{2A_*-A}}{\sqrt{e^{4A}-e^{4A_*}}} = 2\int_{\phi_*}^{0} \dd\phi \frac{e^{2A_*-A}}{\sqrt{e^{4A}-e^{4A_*}}}\parent{\frac{\dd\phi}{\dd\rho}}^{-1}\,.
\end{equation}

Finally, note that a change of the boundary condition $\delta x_1(\infty) = \delta\ell$ causes a change in $S_{q\overline{q}}$ proportional to the conjugate momentum
\begin{equation}
\delta S_{q\overline{q}} = \int^\infty \dd\rho\ \pi_x\, \delta x_1' =   \frac{T}{2\pi\alpha'} \, p \, \delta\ell\,
\end{equation}
in such a way that the change in the potential between the quarks reads
\begin{equation}
	\delta V_{q\overline{q}} = -\frac{1}{2\pi\alpha'}\,p\, \delta \ell
\end{equation}
and the force felt by them is
\begin{equation}
F_{q\overline{q}} = -\frac{\delta V_{q\overline{q}}}{\delta\ell} =  \frac{1}{2\pi\alpha'}\, e^{2A_*}\,.
\end{equation}

\bibliographystyle{JHEP}
\bibliography{refscCFT}

\end{document}